\documentclass[journal=jctcce,manuscript=article]{achemso} 
\usepackage[T1]{fontenc} 
\usepackage[version=3]{mhchem} 
\usepackage[dvipsnames]{xcolor}
\usepackage{graphicx,subcaption}
\usepackage{amssymb}
\usepackage{overpic}
\usepackage{soul}
\usepackage{bm}
\usepackage{graphicx}

\newif\ifhighlightchanges
\highlightchangesfalse

\definecolor{darkblue}{HTML}{003D6D}

\newcommand{\bR}{{\bm R}}
\newcommand{\br}{{\bm r}}
\newcommand{\bP}{{\bm P}}

\newcommand{\bp}{{\bm p}}

\newcommand{\bd}{{\bm d}}
\newcommand{\bG}{{\bm G}}
\newcommand{\bM}{{\bm M}}

\newcommand{\bX}{{\bm X}}
\newcommand{\bGamma}{{\bm \Gamma}}

\newcommand{\hH}{\hat H}

\newcommand{\hp}{\hat p}
\newcommand{\hL}{\hat L}

\newcommand{\hbp}{{\hat{\bm p}}}
\newcommand{\hbP}{{\hat{\bm P}}}
\newcommand{\hbr}{{\hat{\bm r}}}

\newcommand{\hbR}{{\hat{\bm R}}}
\newcommand{\bB}{{\bm B}}
\newcommand{\bA}{{\bm A}}

\newcommand{\hbX}{{\hat{\bm X}}}

\newcommand{\bra}[1]{\left< #1\right|}
\newcommand{\ket}[1]{\left|#1 \right>}
\newcommand{\pp}[2]{\frac{\partial {#1}}{\partial {#2}}}
\newcommand{\hbm}[1]{\hat{\bm{#1}}}
\usepackage{caption}

\title{A Phase-Space Electronic Hamiltonian for Molecules in a Static Magnetic Field I: Conservation of Total Pseudomomentum and Angular Momentum}

\usepackage{geometry}
\author{Mansi Bhati}
\affiliation{Department of Chemistry, Princeton University, Princeton, NJ USA}
\author{Zhen Tao}
\affiliation{Department of Chemistry, Princeton University, Princeton, NJ USA}
\author{Xuezhi Bian}
\affiliation{Department of Chemistry, Princeton University, Princeton, NJ USA}
\author{Jonathan Rawlinson}
\affiliation{Department of Mathematics, Nottingham Trent University, Nottingham, UK}
\author{Robert Littlejohn}
\affiliation{Department of Physics, University of California, Berkeley, California 94720, USA}
\author{Joseph E. Subotnik}
\email{js8441@princeton.edu}
\affiliation{Department of Chemistry, Princeton University, Princeton, NJ USA}

\begin{document}

\maketitle

\begin{abstract}
We develop a phase-space electronic structure theory of molecules in magnetic fields. For a system of electrons in a magnetic field with vector potential $\bA(\hbr)$, the usual Born-Oppenheimer Hamiltonian  is
the sum of the nuclear kinetic energy and the electronic Hamiltonian, $\frac{(\bP - q\bA(\bX) )^2}{2M} + \hat{H}_{e}(\bX)$ (where $q$ is a nuclear charge).  To include the effects of coupled nuclear-electron motion in the presence of magnetic field, we propose that the proper phase-space electronic structure Hamiltonian will be of the form $\frac{(\bP - q^{\textit{eff}}\bA(\bX) - e\hat{\bGamma})^2}{2M} + \hat{H}_{e}(\bX)$. Here,  $q^{\textit{eff}}$ represents the {\em screened} nuclear charges and the $\hat{\bGamma}$ term captures the local pseudomomentum of the electrons. This form reproduces exactly the energy levels for a hydrogen atom in a magnetic field; moreover, single-surface dynamics along the eigenstates  is guaranteed to conserve both the  total pseudomomentum as well as the total angular momentum in the direction of the magnetic field. This Hamiltonian form can be immediately implemented within modern electronic structure packages (where the electronic orbitals will now depend on nuclear position ($\bX$) and nuclear momentum ($\bP$)). One can expect to find novel beyond Born-Oppenheimer magnetic field effects for strong enough fields and/or nonadiabatic systems.
\end{abstract}

\section{Introduction: Dynamics in Magnetic Fields}

Understanding the effect of magnetic fields in chemistry is critical both for characterizing molecules as well as for promoting magnetic field effects and magnetic field dependent chemical reactions more generally.  On the one hand, nowadays it is routine to use large magnetic fields to confirm the molecular identities of a given reaction through NMR\cite{nmrbook}. On the other hand, even though magnetic fields only modestly alter the energy levels of molecular systems (especially the levels of spin states), magnetic field effects on reaction kinetics and product yields are well-known\cite{steiner1989,rodgers2009magnetic}. Interestingly, even magnetic fields as weak as the earth's magnetic field can impact chemical processes in biological systems, e.g. bird magnetoreception \cite{pnas_magnetoreception}. At present, one standing hypothesis is that birds sense magnetic fields through a radical pair mechanism, whereby a slight magnetic field can disrupt the coherence of a singlet-triplet entangled state.\cite{radicalpair,spinweakmagneticfield}
More generally, when a strong magnetic field is applied, it is is known that $(i)$ electrons can become increasingly localized near atomic nuclei leading to an increased binding energy \cite{stopkowicz2015coupled} and $(ii)$ electron motion can also increase the screening of magnetic fields leading to nuclear bond length contraction\cite{cederbaum1990}.  Large magnetic fields can also increase rotational and vibrational couplings\cite{brigham1987,cederbaum1989,lai2001matter}in molecular systems. Predicting such behavior realistically for large molecules is an important goal for chemistry, as the study of magnetic field effects is a rich area of research with many features still not fully explored.

Now, unfortunately, as noted by many chemists already\cite{cederbaum1988}, developing predictive molecular chemistry in a magnetic field $\bB$ is difficult because the effect of electronic momentum becomes more and more important for large $\bB$ fields.  To understand why this is so,  note that magnetic fields deflect moving charges (both nuclei and electrons), but within the Born-Oppenheimer (BO) approximation, the nuclei are frozen on the time scale of the electrons, so that these two fundamental particles are not treated equivalently as far as the effect of $\bB$\cite{yin1994magnetic}. More precisely, when one solves the electronic Schrodinger equation within the BO approximation, one freezes the nuclei and computes the effect of the magnetic field only on the electrons; mathematically, the electronic stationary states depend on nuclear position $\bX$ and magnetic field $\bB$.  Thereafter, if one propagates the nuclear dynamics with momentum $\bm{\Pi}$ along a BO surface, the nuclei will respond to the magnetic field, but the electrons do not experience any additional magnetic field from the fact that the nuclei are moving -- after all, as just described, the electronic eigenstates of the Schrodinger equation were solved with no knowledge of any nuclear momentum \cite{yin1994magnetic}. As a result of this incomplete calculation, at the end of the day, the electrons and nuclei experience the same magnetic field in different ways and at different levels of approximation, and these fields are not fully compatible with each other.  Thus, in the literature, one often reads that nuclei get deflected under a magnetic field as if they were ``bare nuclei''\cite{Schmelcher_1988,yin1994magnetic} and that a lack of electron screening leads to incorrect nuclear charges for dissociated atoms\cite{cederbaum1988}. 
 
For all of the reasons above, there is a clear need to go beyond the BO approximation when modeling dynamics in  magnetic fields. Note, however, that the present BO failures are not the usual BO problems that interest photochemists, i.e,. the presence of conical intersections and/or hops between surfaces that can be treated with Tully's fewest switches surface hopping (FSSH)\cite{hynes:ci_fssh:review,tullysurfacehopping}. Instead, at bottom, the essential problem that is laid bare by a magnetic field is the fact that classical dynamics along BO surfaces do not properly account for electronic momentum\cite{coraline_pssh} (which is usually a small effect because electronic momentum is so much smaller than nuclear momentum). In fact, this failure of the BO approximation can be understood without invoking a magnetic field at all. 
To that end, consider a hydrogen atom initially translating  in space with position $\bX(t)$ (and suppose $\dot{\bX} = {\bm V}_H$ is a constant) with the electron in a $1s$ state. According to the BO approximation, if we assume the nuclei are classical, the electronic state of the system is simply 
\begin{eqnarray}
    \Psi(\br,t) = \phi_{1s}(\br-\bX(t)) = \frac{1}{\sqrt{\pi}a_0^{3/2}} \exp\left(\frac{|\br-\bX(t)|}{a_0}\right),
\end{eqnarray}
where $a_0$ is Bohr radius.
Clearly, the electronic momentum must be $\left<\hat{\bp}_e\right> = m_e {\bm V}_H$. 
However, if one calculates the electronic momentum, the result is zero: $\left< \Psi \middle| \hat{\bp}_e \middle| \Psi \right> = 0$.
Now, as demonstrated in Ref.\citenum{Littlejohn2023},  within BO framework, the putative nuclear momentum actually represents the total nuclear and electronic momentum  per se and so, on the one hand, it is true that the BO approximation does conserve a nominal momentum. Nevertheless,  on the other hand, because the electronic momentum is  zero within the BO approximation, the value of that  momentum conservation is limited for a system of electrons and nuclei\cite{Bian2023}; after all, if one were to include a nonzero electronic momentum, then the total BO momentum would not be conserved.  In short, one would like a framework that both includes nonzero electronic momentum and also conserves total momentum.

Now, one way to approximately account for the electronic momentum classically and to restore momentum conservation is through a Berry force\cite{mead1979determination,berry1984quantal,Bian2023}. Let $\hH_{e}$ be the electronic Hamiltonian for a collection of electrons surrounded by some fixed nuclei, altogether in a magnetic field.
If one diagonalizes such a Hamiltonian:
\begin{eqnarray}
    \hat{H}_{e}(\bm{X})|\Phi_u(\bm{X})\rangle &=& E_u(\bm{X})|\Phi_u(\bm{X})\rangle
\end{eqnarray}
the wavefunctions  $\Phi_u(\br,\bX) = \left< \br \middle | \Phi_u(\bX) \right>$  will be complex valued, and the Berry force on state $u$ is then defined as: 
\begin{eqnarray}
    F_u^{Berry} &=& i\hbar (\nabla_n \times \bm{d}_{uu})\cdot \dot{\bX}
\end{eqnarray}
where $\bm{d}_{uu} = \left< \Phi_u \middle| \nabla \Phi_u \right> $. As we have pointed out \cite{Bian2023} for a system of spins in the absence of a magnetic field, including a Berry force restores the conservation of linear and angular momentum if one includes a nonzero electronic momentum; moreover, as Helgaker {\em et al} have pointed out, in the presence of a magnetic field, including a Berry force also restores the total pseudomomentum\cite{magnetictrans}. 

Unfortunately, however, a naive evaluation of  a Berry Force for adiabatic dynamics can be computationally expensive (though for the case of a GHF ground state,  the calculation is not prohibitive\cite{Culpitt2022}). Moreover, in the presence of electronic degeneracy, e.g. the doublet considered in Ref. \citenum{Bian2023},  the notion of a Berry force in some ways becomes arbitrary because one mostly picks one electronic state out of a degenerate pair. Finally, if one seeks to model photochemical and/or electrochemical dynamics with curve crossings, the possibility of electronic degeneracy renders the surface-hopping procedure ill-defined\cite{Bian2021}. For all of these conceptual and theoretical reasons, there are strong reasons to develop alternatives to Berry forces  when propagating adiabatic dynamics in a magnetic field. \cite{citation-key}

In a set of recent papers\cite{coraline_pssh,tian_erf, tao2024basis}, we have argued that, in the absence of a magnetic field, an alternative approach to both include electronic momentum and  conserve the total momentum  within a nearly BO framework is to use a phase-space electronic Hamiltonian, for which the electronic Hamiltonian and electronic energies are parameterized by both $\bX$ and $\bP$, 
\begin{eqnarray}\label{eq:hps}
  \hat{H}_{PS}(\bX,\bP)\ket{\Phi^{PS}_u(\bX,\bP)} = E_u^{PS}(\bX,\bP) \ket{\Phi^{PS}_u(\bX,\bP)}
\end{eqnarray} 
Such phase-space Hamiltonians can be derived heuristically by including an approximation of the derivative coupling in the electronic Hamiltonian\cite{duston2024phase}.
Now, after diagonalization, the phase-space eigenvalue has a kinetic energy and a potential energy $\epsilon_u(\bX)$ (as in BO theory) but also a mixed energy $f_{u}(\bX,\bP)$(that is a non-separable function of $\bX$ and $\bP$ which is absent in BO theory):
\begin{eqnarray}
   E_u^{PS}(\bX,\bP) = \bP \cdot \frac{1}{2\bM} \cdot \bP  + \epsilon_u(\bX) + f_{u}(\bX,\bP) 
\end{eqnarray} 
so that Hamilton's equations take the form: 
\begin{eqnarray}
    \dot{\bX} &=& \frac{\partial E_{u}^{PS}(\bX,\bP)}{\partial \bP} = \frac{\bP}{\bM} + \frac{\partial f_{u}(\bX,\bP)}{\partial \bP}
    \\
    \dot{\bP} &=& - \frac{\partial E_{u}^{PS}(\bX,\bP)}{\partial \bX} = -\frac{\partial \epsilon_u(\bX)}{\partial \bX} -\frac{\partial f_{u}(\bX,\bP)}{\partial \bX}\label{eq:pdotPS}
\end{eqnarray}
Clearly, nuclear dynamics along such phase-space adiabats incorporate some pseudomagnetic field effects (since $\dot{\bX} \ne \frac{\bP}{\bM}$). 
Moreover, as demonstrated in Refs. \citenum{wu2024linear, coraline_pssh}, if one uses a proper phase-space Hamiltonian, one is guaranteed to conserve the linear and angular momentum in the absence of a magnetic field (and without ever needing to compute a  Berry force).  For these reasons, at present, a great deal of research in our group is now developing a surface hopping algorithm using these phase-space electronic states, a so-called phase-space surface hopping dynamics algorithm.\cite{shenvi:2009:jcp_pssh,bian2024pssh}

In what follows below, our goal will be to extend the results of Refs. \citenum{tian_erf, coraline_pssh,tao2024basis} so as to treat molecules in the presence of a static magnetic field of arbitrary strength (here, in the $z$ direction) with phase-space electronic Hamiltonians. From a theoretical point of view, given that magnetic fields break time-reversal symmetry, use of a phase-space electronic Hamiltonian seems like a very natural way of solving such a problem -- since a $\bP-$dependent Hamiltonian also breaks time-reversal symmetry. In developing the appropriate model, we will insist on several features: $(i)$  the model must conserve the total pseudomomentum of the system, $(ii)$ the model must conserve the angular momentum in the direction of the magnetic field; $(iii)$ the model must conserve the total energy, $(iv)$ the Hamiltonian must also be gauge-invariant and $(v)$ the Hamiltonian eigenspectrum must  be exact for the hydrogen atom.  Thus far, we have found one such meaningful Hamiltonian (reported below), and we expect that running dynamics along the resulting phase-space adiabatic states will yield important dynamical corrections (that are not necessarily  captured by a Berry force\cite{peters2021ab,Bian2023}, as Berry force is zero for real electronic wavefunctions in the absence of magnetic field). 
As a side note, in the future, it will be interesting to compare this current approach with the exact factorization pioneered by Gross, Abedi, Maitra, Agostini and others\cite{PhysRevLett.105.123002,scherrer2015nuclear}, given recent results demonstrating how the latter allows for exchange for momentum between subsystems\cite{PhysRevLett.128.113001}.

An outline of this paper is as follows. In Sec. \ref{sec_intro_mag_field}, we review the Hamiltonian for a collection of nuclei and electrons in the presence of magnetic fields, establish the standard  Born-Oppenheimer framework, review key results from this approach, and highlight the usual semi-classical approaches there. In Sec. \ref{semi_class_ham_PS},
we then introduce the relevant phase-space electronic Hamiltonian (with a new Hamiltonian term ${\bm {\hat{\Gamma}} \cdot \bm P}$)
and derive the subsequent equations of motions. In Sec. \ref{psproperties}, we then list several appealing attributes of this phase space approach, namely gauge invariance to the magnetic origin,  translational and rotational energy independence, and conservation of pseudomomentum and angular momentum in a magnetic field. Finally, in Sec.\ref{ETF-ERF}, we delineate and interpret the final proposed form for the so called electron translation factors (ETFs) and electron rotation factors (ERFs) that are essential for building a phase-space electronic structure Hamiltonian as relevant to dynamics in a magnetic field; in particular, the final form for the $\hat{\bm{\Gamma}}$ term is given in Eqs. \ref{eq:etf_trans} and \ref{eq:final_erf}. In Sec. \ref{results}, we demonstrate that our proposed phase-space Hamiltonian agrees with the exact Hamiltonian for the hydrogen atom. We  conclude in Sec.\ref{conclude}.  We emphasize that, in this paper, the final results are in a complete phase space theory independent of any basis; for a description of how to implement the current phase space theory within an atomic orbital basis, please see the companion Paper II \cite{bhati2024paper2}.

With so much ground to cover, it will be essential to establish clear notation. Our notation will be as follows:
\begin{itemize}
    \item Lower case $\br$ denotes electronic coordinates, capital case $\bX$ denotes nuclear coordinates; more generally, capital letters are used for nuclei and small letters are for electrons.
    \item an ``e'' subscript denotes an electronic observable and a ``n'' subscript denotes a nuclear observable.
    \item The charge of an electron is denoted $-e$ (i.e., we fix $e >0$). The charge of nucleus I is $Q_Ie$. 
    \item $\left\{\alpha, \beta, \gamma, \delta\right\}$ index the $x,y,z$ Cartesian directions.
    \item  ${I,J,...}$ indexes nuclear centers and ${i,j,...}$ indexes electronic centers.
    \item $\bm{G}$ denotes the gauge origin.
    \item $\bB$ denotes a magnetic field corresponding to vector potential $\bm{A}$.
    \item Bold indicates three-dimensional quantities, e.g. $\br$. 
    \item Hats indicate operators (either nuclear or electronic), e.g. $\hat{H}$. 
    \item $u,v$ index adiabatic states
\end{itemize}

\section{Standard Description of a Molecule (or Molecules) in a Magnetic Field}\label{sec_intro_mag_field}

Before discussing a phase approach to dynamics in a magnetic field, for the sake of completeness, let us briefly  review the standard theory for a molecule in a magnetic field. 

\subsection{Fully Quantum Description}

The total non-relativistic Hamiltonian for a set of electrons and nuclei in the presence of an external homogeneous magnetic field can be written as sum of the kinetic energy of nuclei and electrons plus the potential energy of their interactions:
\begin{eqnarray}
    \hat{H}_{mol} &=& \hat{T}_{n} + \hat{H}_e \label{eq:molH}\\
    \hat{H}_e &=& \hat{T}_{e} +\hat{V}_{nn}+\hat{V}_{ne}  + \hat{V}_{ee} \label{eq:Helec} 
\end{eqnarray}
The nuclear kinetic energy $\hat{T}_{n}$ is given by,
\begin{equation}
    \hat{T}_{n} = \sum_{I=1}^{N_{n}} \frac{\hat{\bm{\Pi}}_I^2}{2M_I}= \sum_{I=1}^{N_{n}} \frac{({\mathbf{\hat{P}_I}} - eQ_I{\mathbf{A(\hat{X}_I)}})^2}{2M_I}\label{eq:Tnuc}
\end{equation}
where $\mathbf{A(\hat{X}_I)}$ is the external magnetic vector potential and we define $e>0$ (so that the electronic charge is $-e$). Similarly, the electronic kinetic energy $\hat{T}_{e}$ is,
\begin{equation}
        \hat{T}_{e} = \sum_{i=1}^{N_{e}} \frac{\hat{\bm \pi}_i^2}{2m_e}= \frac{1}{2m_e}\sum_{i=1}^{N_{e}} (\hat{\mathbf{p}}_i + e\mathbf{{A}(\hat{r}}_i))^2\label{eq:kinetic_ele}
\end{equation}
The potential energy is given by $\hat{V}_{tot}$:
\begin{eqnarray}
    \hat{V}_{tot} &=& \hat{V}_{nn}+ \hat{V}_{ee} + \hat{V}_{ne}\\
    &=& \frac{1}{2}\sum_{I,J\neq I}^{N_{n}} \frac{Q_IQ_J e^2}{4\pi \epsilon_0 |\hbX_I-\hbX_J|} + \frac{1}{2}\sum_{i,j\neq i}^{N_{e}} \frac{e^2}{4\pi \epsilon_0 |\hbr_i-\hbr_j|}- \sum_{i=1}^{N_{e}}\sum_{I=1}^{N_{n}} \frac{Q_I e^2}{4\pi \epsilon_0 |\hbr_i-\hbX_I|}\label{eq:Vall}
\end{eqnarray}
When working with uniform magnetic fields, there is effectively a standard form for the vector potential ${\bm A}$ that is broadly used in the literature:
\begin{eqnarray}\label{eqn:Aw/oG}
     \mathbf{A(\hbr)} = \frac{1}{2}\mathbf{B \times \hbr}
\end{eqnarray}
This form satisfies the the standard  Coulomb gauge (i.e., $\nabla \cdot \mathbf{A(\hat{r})} = 0$ ) and is consistent with a magnetic field $\bm B$:
\begin{align}
\label{eq:crossAisB}
    \nabla \times \mathbf{A(\hbr)} = \mathbf{B},
\end{align}
Of course, this potential is not unique and we can define a new potential $\mathbf{A'(\hbr)}= \mathbf{A(\hbr)} + \nabla \mathbf{f(\hbr)}$, which would also satisfy Eq. \ref{eq:crossAisB} above. In practice, the simplest choice of $\mathbf{f(\hbr)}$ would be a linear function,  which is equivalent to shifting gauge origin for the potential.
%
For this reason, below we will define our vector potential to be of the form:
\begin{equation}\label{eqn:vector_potential}
    \mathbf{A(\hat{r})} = \frac{1}{2}\mathbf{B \times (\hat{r}-G)} \; \; \; \; \mbox{(for electrons)}
\end{equation}
and a similar equation for $\mathbf {A(\hbX)}$  for the nuclei.
Here, we have defined $\mathbf{G}$ as the so-called gauge origin and, in what follows, it will be crucial to identify physics that is independent of this choice of $\mathbf{G}$. As a sidenote, henceforward, we will assume that ${\bm B}$ points in the $z$-direction.

Finally, before concluding this introduction, we note that in the presence of a magnetic field, the notion of momentum becomes complicated because there are actually three 
different momentum vectors that can be defined for nuclei and electrons: the standard canonical momentum (that is conjugate to position), the kinetic momentum  (that enters the kinetic energy expression), and the pseudomomentum (which may be unfamiliar to the reader).  For the sake of completeness, let us now define these quantities for a single nucleus and/or electron: 
\begin{center}
\label{table:1}
\begin{tabular}{|c|c|c|c|}
\hline
& Nuclei & Electrons \\
\hline
Canonical &  $\begin{matrix}\hat{P}_n^\alpha &=& -i\hbar \partial/\partial X_\alpha\end{matrix}$ &  $\begin{matrix}\hp_e^\alpha &=& -i\hbar\partial/\partial r_\alpha\end{matrix}$ \\
\hline
Kinetic & $\begin{matrix}\hat{\Pi}_n^\alpha &=&\hat{P}_n^\alpha - \frac{Qe}{2} (\bB \times (\hat{\bX}- \bG))_\alpha\end{matrix}$ & $\begin{matrix}\hat{\pi}_e^\alpha &=&\hat{p}_e^\alpha + \frac{e}{2}(\bB \times (\hat{\br}-\bG))_\alpha\end{matrix}$\\
\hline
Pseudo &  
$\begin{matrix}
\hat K_n^\alpha &=& \hat{ \Pi}_n^\alpha + {Qe} (\bB \times (\hat \bX-\bG))_\alpha \\ & =& \hat P_n^\alpha + \frac{Qe}{2} (\bB \times (\hat \bX-\bG))_\alpha
\end{matrix}$ &
$\begin{matrix}
\hat k_e^\alpha &=&\hat{\pi}_e^\alpha - e (\bB \times (\hbr-\bG))_\alpha\\ &=& \hat p_e^\alpha - \frac{e}{2} (\bB \times (\hbr-\bG))_\alpha
\end{matrix}$ \\
\hline
\end{tabular}
\captionof{table}{The three different momenta: canonical, kinetic, and pseudo.}
\end{center}

Obviously, for conserved quantities, one will need to sum over electrons and nuclei (see below). As shown in Appendix \ref{conserve_proof}, in general, for a set of charges (nuclei and electrons) and a magnetic field in the $z$ direction, there are five conserved quantities corresponding to the following five operators that commute with $\hat H_{mol}$:
\begin{enumerate}
    \item $\hat{ K}_x,\hat{ K}_y,\hat{ K}_z$, the  total pseudomomentum in the $x, y$ and $z$ directions, (summed over all particles)
    \item  $\hat{ L}_z$, the angular momentum in the $z$ direction (summed over all particles)
    \item Obviously, $\hat H_{mol}$ corresponding to $E = \left< \hat H_{mol} \right> $, the total energy.
\end{enumerate}

For our purposes below, it will be helpful to introduce one last quantity, the pseudomomentum of an electron relative to a nuclear coordinate:
\begin{eqnarray}
    \hat{\bm{k}}_I  &\equiv& \hat{\bm{k}}_e +  e\bm{B}\times (\hat{\bm{X}}_I-\bG) \\
    &=& \bm{\hat{\pi}}_e -e\bm{B}\times (\bm{\hat{r}}-\hat{\bm{X}}_I) \\&=& \bm{\hat{p}}_e - \frac{e}{2}\bm{B}\times (\hbr-\bG)+  e\bm{B}\times (\hat{\bm{X}}_I-\bG)\label{eq:pseudoAdefn}
\end{eqnarray}

\subsection{Born-Oppenheimer Framework}\label{BO_framework}
Diagonalizing the Hamiltonian in Eq. \ref{eq:molH} above is daunting and usually impossible. To that end, 
within a Born-Oppenheimer framework, one expresses the total molecular wavefunction in a basis of electronic eigenstates $\left\{ \ket{\Phi(\bm{X})} \right\}$ that depend parametrically on nuclear coordinates. 
Mathematically, we diagonalize the operator $ \hat{H}_{e}$ from Eq. \ref{eq:Helec} above,
\begin{eqnarray}
    \hat{H}_{e}\ket{\Phi_u(\bm{X})} = \Big(\hat{T}_{e} + \hat{V}_{ee}+ \hat{V}_{ne}+\hat{V}_{nn}\Big)\ket{\Phi_u(\bm{X})} = E^{BO}_u (\bm{X}) \ket{\Phi_u(\bm{X})}\label{eq:He}
\end{eqnarray}
in order to generate $\ket{\Phi_u(\bm{X})}$.

In this basis, the total molecular Hamiltonian becomes:
\begin{eqnarray}
    \hat{H}_{mol} &=& \sum_{Iuvw} \frac{1}{2M_I} \ket{\Phi_u} \left( \bm{\hat{\Pi}}_I \delta_{uv} - i\hbar \bm{{d}}_{uv}^I \right)\cdot
    \left( \bm{\hat{\Pi}}_I \delta_{vw} - i \hbar \bm{{d}}_{vw}^I \right)\bra{\Phi_w} 
    \nonumber \\
    &+& \sum_v E_{vv}({\bm \hat{X}})\ket{\Phi_v}\bra{\Phi_v}
    \label{eq:HBO_d}
    \end{eqnarray}
where (according to Table \ref{table:1}), 
    \begin{eqnarray}
\bm{\hat{\Pi}}_I = \mathbf{\hat{P}_I} - eQ_I{\mathbf{A(\hat{X}_I)}} 
\end{eqnarray}
As is well known, in the BO framework, the derivative coupling $\hat{\bm{d}}$ emerges as the primary electron-phonon coupling and is of the form:
\begin{eqnarray}
    \bm{{d}}_{uv}^I(\bG) &=& \left< \Phi_u \middle| \frac{\partial}{\partial \bX_I} \Phi_v\right>  \\
    &= & \frac{\left< \Phi_u \middle|  \partial \hat{H}_e/\partial \bX_I \middle|\Phi_v \right> }{E_v - E_u}\label{eq:ddef}
\end{eqnarray}
where the second equality follows from the Hellman-Feynman theorem. Note that $\hat{\bm{d}}$ depends on the magnetic origin $\bG$ because $\ket{\Phi_u}$ depends on $\bG$.

\subsection{The Derivative Coupling Sum-Rule}
A crucial  sum-rule applies to the derivative coupling. In the absence of a magnetic field, one can show that, by translational invariance, the electronic Hamiltonian commutes with the total momentum, 
\begin{eqnarray}\label{eq:Hetrans1}
[\hbP + \hbp, \hat{H}_e] = 0
\end{eqnarray}
Here, $\hbP = \sum_I \hbP_I$ is the total nuclear momentum  and  $\hbp = \sum_i \hat{\bm{p}}_{i}$ is the total electronic momentum (summed over all electrons).  
The equality in Eq. \ref{eq:Hetrans1} reflects the fact that the Hamiltonian is unchanged if one moves the electrons and nuclei together. One can rewrite this equation as:
\begin{eqnarray}\label{eq:Hetrans2}
    -i\hbar\sum_I\frac{\partial \hH_e}{\partial \bX_I} = [\hH_e,\hbp]
\end{eqnarray}
From this relationship, it follows from Eq. \ref{eq:ddef} that (for $u \ne v$)
\begin{eqnarray}
\label{sumrule_nomag}
    i\hbar\sum_I {\bm{d}}^I_{uv} = {\bm{p}}_{uv}
\end{eqnarray}
In words, if we sum the derivative coupling over all the nuclei, the result is the electronic momentum (divided by $i\hbar$). More generally, Eq. \ref{sumrule_nomag} can be viewed as a phase convention for the adiabatic states, where we assert that  
\begin{eqnarray}
\left(\sum_I -i\hbar \frac{\partial}{\partial \bX_I} + \hbp\right) \ket{\Phi_v} &=& 0 \label{eq:phaseduv}
\end{eqnarray}
This choice is equivalent to insisting that the electronic wavefunction is expressed relative to the nuclear coordinates (i.e. an adiabat is of the form $\Phi_v(\br - \bX_1, \br-\bX_2, \ldots)$).

Now, in the context of a problem with a magnetic field, the sum-rule is slightly different as the magnetic field breaks translational symmetry. To derive the proper relationship, note that, if one translates the electrons, the nuclei, and the magnetic origin $\bG$, the Hamiltonian must still be unchanged. 
\begin{eqnarray}
-i\hbar \Bigg[\sum_I\frac{\partial}{\partial \bX_I} +\sum_i\frac{\partial}{\partial \br_i} + \frac{\partial}{\partial \bG},\hat{H}_e \Bigg]&=& 0\label{eq:sum_b}
\end{eqnarray}
Mathematically, for $\hat{H}_e$ in Eq. \ref{eq:He}, this invariance implies:
\begin{align}
[-i\hbar \sum_I\frac{\partial}{\partial \bX_I},\hat{H}_e] + [\hbp -i\hbar \frac{\partial}{\partial \bG}, \hat{H}_e]&= 0\\
[-i\hbar \sum_I\frac{\partial}{\partial \bX_I},\hat{H}_e] + [\hbp, \hat{H}_e] + i\hbar(2\hat{\bm{\pi}})\frac{\partial }{\partial \bG}\Big(\frac{e}{2}\bB\times \bG\Big)&= 0\\
[-i\hbar \sum_I\frac{\partial}{\partial \bX_I},\hat{H}_e] + [\hbp, \hat{H}_e] - i\hbar(2\hat{\bm{\pi}})\frac{\partial }{\partial \br}\Big(\frac{e}{2}\bB\times \hbr\Big)&= 0\\
[-i\hbar \sum_I\frac{\partial}{\partial \bX_I},\hat{H}_e] + [\hbp -\frac{e}{2}(\bB \times \hbr ), \hat{H}_e] &= 0 \label{eq:kcommute1} 
\end{align}
At this point, we can invoke the definition of the pseudomomentum $\bm{k}$ in Table \ref{table:1}:
\begin{align}
[-i\hbar \sum_I\frac{\partial}{\partial \bX_I},\hat{H}_e] + [\hat{\bm k}, \hat{H}_e] &= 0 \label{eq:kcommute2}
\end{align}
From this relationship and Eq. \ref{eq:ddef},
the result in a magnetic field is that (for $u \ne v$) 
\begin{eqnarray}
\label{sumrule_yesmag}
    i\hbar\sum_I \bd^I_{uv} = {\bm k}_{uv} 
\end{eqnarray}
In words, if we sum the derivative coupling over all the nuclei in the presence of a magnetic field, the result is the electronic pseudomomentum (up to a constant factor of $i \hbar$). In principle one might like to extend Eq. \ref{sumrule_yesmag} to form a more general analogue of Eq. 
\ref{eq:phaseduv} above (only now replacing $\hbp$ with $\hat{
\bm{k}}$ in the presence of a magnetic field):
\begin{eqnarray}
\left(\sum_I -i\hbar \frac{\partial}{\partial \bX_I} + \hat{\bm{k}}\right) \ket{\Phi_v} &\stackrel{?}{=}& 0 \label{eq:phasekuv}
\end{eqnarray}
Unfortunately, however,  we note that such a phase convention is not possible in general in three dimensional space because, unlike for momentum, the different components of the electronic pseudomomentum do not commute with each other (so that the resulting phases would be path-dependent).

\subsection{Gauge Invariance of the Born-Oppenheimer Electronic \\Hamiltonian}

One of the most important properties of the Born-Oppenheimer electronic Hamiltonian (Eq. \ref{eq:He}) is gauge invariance, namely the invariance of the electronic energy to the position of the magnetic origin. Mathematically, suppose we have diagonalized the BO Hamiltonian with magnetic origin $\bG$: 
\begin{eqnarray}
& & \hat{H}_{BO}(\bX,\bB,\bG) \Psi(\br;\bX,\bG) =   E_{BO}(\bX,\bB,\bG) \Psi(\br;\bX,\bG)
\end{eqnarray}
If we change the magnetic origin, $\bG \rightarrow \bG+\Delta$, the wavefunction merely picks up a phase, $\Psi(\br;\bX,\bG + \bm{\Delta}) = \Psi(\br;\bX,\bG) \exp(-\frac{ie}{2\hbar}(\bB \times \bm{\Delta}) \cdot \br)$, which is equivalent to  writing:
\begin{eqnarray}
& & \hat{H}_{BO}(\bX,\bB,\bG+ \bm{\Delta}) \Psi(\br;\bX,\bG)\exp(-\frac{ie}{2\hbar}(\bB \times \bm{\Delta}) \cdot \br) = \nonumber\\
& & \; \; \; \; \; \; \; \; \; \; \; \; \; \; \; \; \; \; \; \; \; \;   E_{BO}(\bX,\bB,\bG) \Psi(\br;\bX,\bG)\exp(-\frac{ie}{2\hbar}(\bB \times \bm{\Delta}) \cdot \br) 
\end{eqnarray}
A change in origin leaves the energy and all physical observables unchanged.

\subsection{The Born-Oppenheimer Framework and the \\Born-Oppenheimer Approximation}\label{semi_class_ham_BO}
The BO {\em framework} stipulates that one solves quantum problems using the basis of electronic eigenstate parameterized by nuclear position; indeed, one can recover the exact quantum answer using such a BO framework. That being said, in order to feasibly make predictions, two {\em approximations} often go hand-in-hand within the BO framework. The first approximation is the assumption of classical nuclei; within this approximation, one replaces the nuclear operator $\hbP$ with the classical scalar variable.  
The second approximation is 
 the BO approximation whereby the ansatz that  the total (nuclear plus electronic) molecular wavefunction can be written as the product of a nuclear $\chi(\bm X)$ and electronic $\Phi(\bm{X},\bm{r})$ (which is usually the ground state, $\Phi_g(\bm{X},\bm{r})$) wavefunction:
 
\begin{eqnarray}
    \Psi(\bm{X},\bm{r}) = \Phi(\bm{X},\bm{r})\chi(\bm{X})
\end{eqnarray}
Note that this ansatz is equivalent to ignoring the derivative coupling in Eq. \ref{eq:HBO_d}:
\begin{eqnarray}
    {\bm d}_{uv} \stackrel{BO }{\longrightarrow} 0
\end{eqnarray}
Classical BO theory then posits that the effective nuclear equation of motion is:
\begin{eqnarray}
 \frac{d\bP}{dt} = - \frac{\partial}{\partial \bX} \left< \Phi \middle|\hat{H}_{mol}\middle|\Phi \right> \label{eq:dPdt} 
\end{eqnarray}
This completes our review of standard BO theory in a magnetic field.

\section{A Phase-Space Electronic Hamiltonian}\label{semi_class_ham_PS}

 The premise of phase-space theory is that one does not need to throw out entirely the derivative coupling in  Eq. \ref{eq:HBO_d}.  Rather, one should seek a nuclear-electron coupling term $\bm{\hat{\Gamma}}$ which approximates the derivative coupling:
\begin{eqnarray}
\bd_{uv}^I \approx  \left<  \Phi_u \middle|\bm{\hat \Gamma}_I \middle|\Phi_v \right> \label{eq:approxgamma}
\end{eqnarray}
Obviously, it is impossible to find a simple one-electron operator $\bm{\hat{\Gamma}}$ that satisfies Eq. \ref{eq:ddef}  exactly, but there are a few properties that can be recovered correctly. 

\subsection{Brief Review Without a Magnetic Field}\label{4cond}

Consider initially the case without a magnetic field, which is considered in Refs. \citenum{coraline_pssh,tao2024basis}. 
When choosing the approximate form of $\hat{\bGamma}$, the most important consideration is that the operator must satisfy the constraint in Eq. \ref{sumrule_nomag}.
After all, such behavior (following Noether's theorem) should be related to momentum conservation.
Indeed, according to Ref. \citenum{tao2024basis}, in order to conserve linear and angular momentum, the $\hat{ \bm \Gamma}$ operator must satisfy the following four conditions:
\begin{align}
    -i\hbar\sum_{I}\hat{\bm \Gamma}_{I} + \hbm{p} &= 0,\label{eq:Gamma_uv1_nb}  \\
    \Big[-i\hbar\sum_{J}\pp{}{\bm{X}_J} + \hbm{p}, \hat{\bm \Gamma}_{I}\Big] &= 0,\label{eq:Gamma_uv2_nb}\\
    -i\hbar\sum_{I}({\bm X}_{I}-\bG) \times \hat{\bm \Gamma}_{I} + (\hbm{r} - 
    \bG) \times \hbm{p}+ \hbm{s} &= 0,\label{eq:Gamma_uv3_nb}\\
     \Big[-i\hbar\sum_{J}\left((\bm{X}_J-\bG) \times\pp{}{\bm{X}_J}\right)_{\gamma} + \left((\hbm{r} - 
    \bG) \times \hbm{p}\right)_{\gamma} + \hat{s}_{\gamma}, \hat{\Gamma}_{I \delta}\Big] 
     &= i\hbar \sum_{\alpha} \epsilon_{\alpha \gamma \delta} \hat{\Gamma}_{I \alpha}\label{eq:Gamma_uv4_nb}
\end{align} 
If we sandwich Eq. \ref{eq:Gamma_uv1_nb} between two different electronic states ($\ket{\Phi_u}$ and $\ket{\Phi_v}$), then Eq. \ref{eq:Gamma_uv1_nb} requires that the $\hat{\bGamma}$ operator satisfy Eq. \ref{sumrule_nomag} (in lieu of $\bd$), which is akin to the phase convention for translation in Eq. \ref{eq:phaseduv} above. 
Similarly, if we were to replace $\hat{\bGamma}_I$ with $\hat{\bd}$ in Eq. \ref{eq:Gamma_uv3_nb}, we would find that this equation is a phase convention dictating that the basis of adiabatic electronic states are written down in relative coordinates (where the electronic coordinates are expressed relative to nuclear coordinates upon rotation).
 Eqs. \ref{eq:Gamma_uv2_nb} and \ref{eq:Gamma_uv4_nb} codify the fact that the $\hat{ \bm \Gamma}$ operators must transform correctly with translations and rotations. 
In the absence of a magnetic field, Ref. \citenum{coraline_pssh, tao2024basis} demonstrates that one meaningful choice of $\hat{ \bm \Gamma}$ is to use electron translation factors (ETFs)\cite{fatehi_etf,schneiderman:1969:pr:etf} and electron rotation factors (ERFs)\cite{tian_erf,athavale:2023:erf} which we will discuss further in Sec. \ref{ETF-ERF}. In such a case, one requires a partition of space characterized by different nuclei, and we usually pick this partition function as:
\begin{eqnarray}
    \hat{\Theta}_I(\hbr,\bm{X}) &=&\frac{ m_{I} e^{-|\hbr-\bm{X}_I|^2/\sigma_I^2}}{\sum_J m_{J}e^{-|\hbr-\bm{X}_J|^2/\sigma_J^2}}\label{eq:thetadefn}
\end{eqnarray}
Here, $\sigma_I$ governs the partition of electrons around atom $I$, and can be parametrized using chemical properties such as the Van der Waals radius or electronegativity. Following the arguments of Ref. \citenum{tao2024basis}, we will show in Sec. \ref{ETF-ERF} how we can use the characteristic function in Eq. \ref{eq:thetadefn} to generate the relevant $\hat{\bGamma}$ operators.

\subsection{$\hat H_{PS}$ in the Presence of a Magnetic Field}\label{hps_magfield}

As discussed above in Sec. \ref{sec_intro_mag_field}, 
in the presence of a magnetic field, the total linear and angular momentum of the system are no longer conserved quantities.  Instead, the pseudomomentum in all three directions and the canonical angular momentum in the direction of the magnetic field are the new conserved quantities.\cite{Greenshields2014,Johnson1983} (For a quick proof, see Appendix \ref{conserve_proof}.) 
Moreover, as shown in Eq. \ref{sumrule_yesmag}, the sum over atoms of the derivative couplings in the presence of a magnetic field is no longer the electronic momentum, but now rather the pseudomomentum. For these reasons, 
 we must expect to replace the conditions for the $\hat{\bGamma}$ operators as set forth in Eqs. \ref{eq:Gamma_uv1_nb} and \ref{eq:Gamma_uv3_nb} above (while keeping the translational and rotational invariance of Eqs. \ref{eq:Gamma_uv2_nb} and \ref{eq:Gamma_uv4_nb}). Given the need to approximate a derivative coupling satisfying Eq. \ref{sumrule_yesmag}, 
 one might propose to replace Eq. \ref{eq:Gamma_uv1_nb} with:
 \begin{eqnarray}
     -i\hbar\sum_{I}{ \hat{\bm \Gamma}}_{I} + \hat{\bm k} &\stackrel{?}{=} 0,\label{eq:nonono}
 \end{eqnarray}
 However, as argued below Eq. \ref{eq:phasekuv} above, the true derivative couplings cannot satisfy such a relationship, and so enforcing such a constraint on the $\hat{\bGamma}$ operators is not prudent. As an alternative, we submit that a better, more meaningful set of  conditions is:
\begin{align}
    -i\hbar\sum_{I}{ \hat{\bm \Gamma}}_{I} + \sum_{I}\hat{\Theta}_I{ \hat{\bm k}_I} &= 0,\label{eq:Gamma_uv1} \\
    -i\hbar\sum_{I}({\bm X}_{I}-\bm G) \times {\hat{\bm \Gamma}}_{I} + \sum_{I}\hat{\Theta}_I (\hat{\bm{r}}-\bm{G})\times {\hat{\bm k}_I} &= 0,\label{eq:Gamma_uv3} 
 \end{align} 
Note that, in Eqs. \ref{eq:Gamma_uv1}-\ref{eq:Gamma_uv3}, 
both
$ \sum_{I}  \hat{\Theta}_I{ \hat{\bm k}_I}$ and $ \sum_{I}  \hat{\Theta}_I (\hat{\bm{r}}-\bm{G})\times{\hat{\bm k}_I}$ are hermitian operators, but that whereas we might have expected to find the electronic pseudomomentum $\hat{\bm k}_e$ (Table \ref{table:1}) in these constraints, instead we now find the relative pseudomomentum $\hat{\bm{k}}_I$ (Eq. \ref{eq:pseudoAdefn}).
While this replacement might at first appear unnatural, such a replacement is in fact necessary to ensure translational invariance. Furthermore, we will show below (see Sec. \ref{justify}) that this replacement can  be rationalized if we reinterpret the effective charge on each atom correctly.
At the end of the day, our ansatz is then to create a phase-space electronic Hamiltonian of the form,
\begin{eqnarray}
\hat{H}_{PS}(\bm X,\bP,\bm G, \bm B)&=&\sum_{I}\frac{1}{2M_I} \left( \bP_I - \frac{1}{2} q^{\textit{eff}}_I(\bX)(\bB\times (\bX_I-\bG))-i\hbar \hbm{\Gamma}_I(\bX) \right)^2\nonumber\\&&+  \hH_{e}(\bX,\bG,\bB)\label{eq:PSH}
\end{eqnarray}
The quantity $q^{\textit{eff}}$
in Eq. \ref{eq:PSH} denotes the  {\em screened} charge on each nucleus. For instance, in the case of a hydrogen atom (one nucleus, one electron), $q^{\textit{eff}}$ = 0.  In the other extreme, for a system of nuclear charges without a single electron, $q^{\textit{eff}} = q_I$. 
We will describe how to calculate 
$q^{\textit{eff}}$
in detail below, in Sec. \ref{qeff_define}. However, for the time being, it is important to note that these effective charges should be functions of the shape of the molecule, invariant to translation or rotation of the molecule. 
As such, these screened charges must satisfy:
\begin{eqnarray}
\label{eq:qeff_inv_trans}
    \sum_{I} \frac{\partial q^{\textit{eff}}_J(\bX)}{\partial X_{I\alpha}}=0
\end{eqnarray}
and 
\begin{eqnarray}
\label{eq:qeff_inv_rot}
\sum_{I \beta \alpha} \epsilon_{z \beta \alpha} X_{I \beta}
    \frac{\partial q^{\textit{eff}}_J(\bX)}{\partial X_{I\alpha}}=0
\end{eqnarray}
Henceforward, to distinguish the nuclear kinetic momentum with $q^{\textit{eff}}_I$ from the raw kinetic momentum, we will define:
\begin{eqnarray}
    \bm{\Pi}^{\textit{eff}}_I&=&\bP_I - \frac{1}{2} q^{\textit{eff}}_I(\bX)(\bB\times (\bX_I -\bG))\label{eq:pieffdefn}
\end{eqnarray}
and the kinetic momentum of nucleus $I$ as:
\begin{eqnarray}
\bm{\Pi}^{\textit{kin}}_I&=&\bP_I - \frac{1}{2} q^{\textit{eff}}_I(\bX)(\bB\times (\bX_I -\bG)) - i\hbar \hbm{\Gamma}_I(\bX)\label{eq:pikindefn} \\
& = & \bm{\Pi}^{\textit{eff}}_I
- i\hbar \hbm{\Gamma}_I(\bX)
\end{eqnarray}


\subsection{Phase-Space Equations of Motion}
For the Hamiltonian in Eq. \ref{eq:PSH}, the  energy is:
\begin{eqnarray}
\label{eq:pssh_e}
    E_{\rm PS}(\bm X,\bP,\bm G, \bm B) &=& \sum_I\frac{(\bm{\Pi}_I^{\textit{eff}})^2}{2M_I} + {V}_{PS}(\bX,\bP,\bm G, \bm B) \\
    {V}_{PS}(\bX,\bP,\bm G, \bm B) & = & 
    \left< \Psi \middle| 
      \hat{H}_{\Gamma}+ \hat{H}_e \middle| \Psi \right>\label{eq:std_e}
\end{eqnarray}
where
\begin{eqnarray}
\hat{V}_{PS} &=& \hat{H}_{\Gamma} + \hat{H}_e \\
\hat{H}_{\Gamma} &=&
      -i\hbar \sum_{I} \frac{\bm{\Pi}_I^{eff}\cdot\bm{\hat{\Gamma}}_{I}}{M_{I}}-\hbar^2\frac{\bm{\hat{\Gamma}}_{I}\cdot \bm{\hat{\Gamma}}_{I}}{2M_I}
      \label{eq:Hgamma}
      \\
\hat{H}_e &=& \frac{\bm{\hat{\pi}}^2}{2m_e} + \hat{V}_{tot}
\end{eqnarray}
In Eq. \ref{eq:pssh_e}, we have separated the kinetic nuclear energy from the phase-space potential energy ${V}_{PS}$, the latter being decomposed as the electronic energy plus a nuclear-electronic coupling term and the coupling term squared in Eq. \ref{eq:std_e}. 

Using Eqs. \ref{eq:pssh_e}-\ref{eq:std_e} above, we can find phase-space equations of motion through Hamilton's principle:
\begin{eqnarray}    \dot{X}_{I\alpha}&=&\left(\frac{\partial E_{PS}}{\partial P_{I\alpha}}\right)_X\\
    &=&\frac{ \Pi^{\textit{eff}}_{I\alpha}}{M_{I}} - i\hbar \left< \Psi \middle|\frac{\hat{\Gamma}_{I\alpha}}{M_{I}}\middle| \Psi \right>\label{eq:xdot}
\end{eqnarray}
\begin{eqnarray}
    \dot{P}_{I\alpha}&=&- \left(\frac{\partial E_{PS}}{\partial X_{I\alpha}}\right)_P  \\
        &=&-\sum_{J\beta} \frac{\Pi^{\textit{eff}}_{J\beta}}{M_J}\frac{\partial }{\partial X_{I\alpha}}\Bigg(- \frac{1}{2} q^{\textit{eff}}_J(\bX)(\bB\times (\bX_J-\bG))_{\beta}\Bigg)_P 
        -\left(\frac{\partial V_{PS}}{\partial X_{I \alpha}}\right)_{P} \\
        &=&-\sum_{J\beta} \frac{\Pi^{\textit{eff}}_{J\beta}}{M_J}\Bigg(- \frac{1}{2} \frac{\partial q^{\textit{eff}}_J(\bX)}{\partial X_{I\alpha}}(\bB\times (\bX_J-\bG))_{\beta}- \frac{1}{2} q^{\textit{eff}}_J(\bX)\frac{\partial (\bB\times (\bX_J-\bG))_{\beta}}{\partial X_{I\alpha}}\Bigg) 
        \nonumber\\&&-\left(\frac{\partial V_{PS}}{\partial X_{I \alpha}}\right)_{P}
        \label{eq:pdot}
\end{eqnarray}
We can further expand the gradient of the phase-space potential energy $V_{PS}$ while keeping $\bP$ constant
into two terms: the first term involves the gradient of the vector potential, and the second term is the gradient of $V_{PS}$ holding $\Pi^{\textit{eff}}$ constant:
\begin{eqnarray}
        \left(\frac{\partial V_{PS}}{\partial X_{I \alpha}}\right)_{P}&=&  - i\hbar \left< \Psi \middle| \sum_{J\beta} \frac{1}{M_J}\frac{\partial }{\partial X_{I\alpha}}\Bigg(- \frac{1}{2} q^{\textit{eff}}_J(\bX)(\bB\times (\bX_J-\bG))_{\beta}\Bigg)_P {\hat \Gamma}_{J\beta}\middle| \Psi \right> \nonumber\\ 
        &&+
\left(\frac{\partial V_{PS}}{\partial X_{I \alpha}}\right)_{\Pi^{\textit{eff}}}\label{eq:Estdgrad0} \\
\label{eq:Estdgrad}
&=&  - i\hbar \left< \Psi \middle| \sum_{J\beta} \frac{1}{M_J}\Big(- \frac{1}{2} \frac{\partial q^{\textit{eff}}_J(\bX)}{\partial X_{I\alpha}} \Big) (\bB\times (\bX_J-\bG))_{\beta} {\hat{\Gamma}}_{J\beta}\middle| \Psi \right>  \nonumber
\\&&-
\nonumber i\hbar \left< \Psi \middle| \sum_{J\beta} \frac{1}{M_J}\Big(- \frac{1}{2} q^{\textit{eff}}_J(\bX)
\Big)
\Bigg(\frac{\partial (\bB\times (\bX_J-\bG))_{\beta} }{\partial X_{I\alpha}}\Bigg)_P {\hat{\Gamma}}_{J\beta}\middle| \Psi \right> \nonumber\\&&+
\left(\frac{\partial V_{PS}}{\partial X_{I \alpha}}\right)_{\Pi^{\textit{eff}}} \\
     \left(\frac{\partial V_{PS}}{\partial X_{I \alpha}}\right)_{\Pi^{\textit{eff}}} & = & 
     \left< \Psi \middle| -\sum_{J\beta}\frac{i\hbar \Pi^{\textit{eff}}_{J\beta}}{M^{J}} \frac{\partial {\hat{\Gamma}}_{J\beta}}{\partial X_{I\alpha}} -\hbar^2 \sum_{J\beta}\frac{\hat{\Gamma}_{J\beta}}{M_J}\frac{\partial {\hat{\Gamma}}_{J\beta}}{\partial X_{I\alpha}}   + \frac{\partial \hat{V}_{tot} }{\partial X_{I\alpha}}\middle| \Psi \right>    \label{eq:gradEpi}
\end{eqnarray}
From the equations of motion  derived above  in addition to the more general rules outlined in Sec. \ref{hps_magfield} (Eqs. \ref{eq:Gamma_uv1}-\ref{eq:Gamma_uv3}, and Eqs. \ref{eq:Gamma_uv2_nb} and \ref{eq:Gamma_uv4_nb}), we can now prove several appealing properties of the electronic phase-space Hamiltonian approach.

\section{Symmetries of the Phase-Space Electronic Hamiltonian} \label{psproperties}
  Before proving any conservation rules, however, a few words are appropriate about the choice of magnetic origin and gauge invariance.

\subsection{Gauge Invariance of the Phase-Space Electronic Hamiltonian}

It is crucial to emphasize that below (Sec. \ref{ETF-ERF}), we will parameterize the $\hat{\bGamma}$ operator 
to be a function of the electronic kinetic momentum and the  position of the electrons relative to the nuclei:

\begin{eqnarray}
    \hat{\bGamma} = \hat{\bGamma}(\hat{\bm{\pi}}, \hat \br - \bX_1, \hat \br-\bX_2, ...)
\end{eqnarray}
In such a way, the total Hamiltonian must be gauge invariant. In other words, if have diagonalized $\hat{H}_{PS}$ for one value of $\bG$,
\begin{eqnarray}
\hat{H}_{PS}(\bX,\bP,\bB,\bG) \Psi(\br;\bX,\bP,\bB,\bG) &=& E_{PS}(\bX,\bP,\bB,\bG) \Psi(\br;\bX,\bP,\bB,\bG)
\end{eqnarray}
it follows that we also know the eigenvalues and eigenvectors for all other values of $\bG$:
\begin{eqnarray}
& & \hat{H}_{PS}(\bX,\bP,\bB,\bG+ \bm{\Delta}) \Psi(\br;\bX,\bP,\bB,\bG)\exp(-\frac{ie}{2\hbar}(\bB \times \bm{\Delta}) \cdot \br) = \nonumber\\
& & \; \; \; \; \; \; \; \; \; \; \; \; \; \; \; \; \; \; \; \; \; \;   E_{PS}(\bX,\bP,\bB,\bG) \Psi(\br;\bX,\bP,\bB,\bG)\exp(-\frac{ie}{2\hbar}(\bB \times \bm{\Delta}) \cdot \br) 
\end{eqnarray}
In other words, 
\begin{eqnarray}
    \Big(\frac{\partial E_{PS}}{\partial\bG}\Big)_{\bm{\Pi}^{\it{eff}}} = 0 \label{eq:eps_g_grad}
\end{eqnarray}
We will discuss gauge invariance in great detail in Paper II \cite{bhati2024paper2}.

.

\subsection{Translational Invariance of $E_{PS}$}\label{trans_inv_eps}
Let us now prove that the phase-space energy is translationally invariant,
\begin{eqnarray}
    \sum_I \Big(\frac{\partial E_{PS}}{\partial \bX_I}\Big)_{\bm{\Pi}^{\textit{eff}}} &=& 0
\end{eqnarray}
According to Eq. \ref{eq:pssh_e}, if we hold $\bm{\Pi}^{\textit{eff}}$ constant, the gradient over the phase-space energy reduces to the gradient of the phase space potential energy $V_{PS}$. Using the Hellmann-Feynman Theorem, and according to Eq. \ref{eq:std_e}, this gradient can be written as:
\begin{eqnarray}
    \sum_I \Big(\frac{\partial V_{PS}}{\partial \bX_I}\Big)_{\bm{\Pi}^{\textit{eff}}} &=& \sum_I \left< \Psi \middle|\Big(\frac{\partial \hat{V}_{PS}}{\partial \bX_I}\Big)_{\bm{\Pi}^{\textit{eff}}}\middle|\Psi \right> \label{eq:destdx}
\end{eqnarray}
In Eq. \ref{eq:destdx}, the electronic kinetic energy and electron-electron repulsion operators do not depend on nuclear coordinates.  The other terms can then be written as:
\begin{eqnarray}
    \sum_I \Big(\frac{\partial V_{PS}}{\partial \bX_I}\Big)_{\bm{\Pi}^{\textit{eff}}} &=& \sum_I \left< \Psi \middle| \frac{\partial \hat{H}_\Gamma}{\partial \bX_I}+ \frac{\partial {V}_{nn}}{\partial \bX_I}+\frac{\partial \hat{V}_{ne}}{\partial \bX_I}\middle|\Psi \right>
\end{eqnarray}
Using Eq. \ref{eq:Gamma_uv2_nb} and the analogue of Eq. \ref{eq:Hetrans2} for $\hat{V}_{ne}$, we can rewrite the gradient in terms of electronic coordinates: 
\begin{eqnarray}
    \sum_I \Big(\frac{\partial V_{PS}}{\partial \bX_I}\Big)_{\bm{\Pi}^{\textit{eff}}}&=&  -\left< \Psi \middle| \frac{\partial \hat{H}_\Gamma}{\partial \br}+\frac{\partial \hat{V}_{ne}}{\partial \br}\middle|\Psi \right> + \sum_I \left< \Psi \middle| \frac{\partial {V}_{nn}}{\partial \bX_I}\middle|\Psi \right> \label{eq:Hstdgrad}
\end{eqnarray}
Now, it is straightforward to prove that $\sum_I  \frac{\partial {V}_{nn}}{\partial \bX_I} = 0$. Furthermore, if we add and subtract a term $\left< \Psi \middle| \frac{\partial \hat{T}_e}{\partial r}\middle|\Psi \right>$ to Eq. \ref{eq:Hstdgrad}, we can identify a  commutator of $\hat{V}_{PS}$:
\begin{eqnarray}
    \sum_I \Big(\frac{\partial V_{PS}}{\partial \bX_I}\Big)_{\bm{\Pi}^{\textit{eff}}}&=&  \frac{1}{i\hbar}\left< \Psi \middle| \Big[\hbp,\hat{V}_{PS}\Big]\middle|\Psi \right>+\left< \Psi\middle| \frac{\partial \hat{T}_e}{\partial \br}\middle|\Psi \right>
\end{eqnarray}

Note that $\left< \Psi \middle| \Big[\hbp,\hat{V}_{PS}\Big]\middle|\Psi \right> = \left< \Psi \middle| \hbp V_{PS}\middle|\Psi \right>-\left< \Psi \middle|V_{PS} \hbp \middle|\Psi \right> =0 $. Furthermore, for $\hat{T}_e$ defined in Eq. \ref{eq:kinetic_ele}, it is straightforward to prove that:
\begin{eqnarray}
    \Big(\frac{\partial \hat{T}_e}{\partial \br}+\frac{\partial \hat{T}_e}{\partial \bG} \Big)_{\bm{\Pi}^{\textit{eff}}} &=& 0
\end{eqnarray}
so that
\begin{eqnarray}
    \sum_I \Big(\frac{\partial V_{PS}}{\partial \bX_I}\Big)_{\Pi^{\textit{eff}}} &=& \left< \Psi \middle| \frac{\partial \hat{T}_e}{\partial \br}\middle|\Psi \right> = -\left<\Psi \middle| \frac{\partial \hat{T}_e}{\partial \bG}\middle|\Psi \right> 
\end{eqnarray}
Finally, noting that for $\hat{V}_{PS}$  in Eq. \ref{eq:std_e}, only the $\hat{T}_e$ operator depends of $\bG$, it follows that
\begin{eqnarray}
\sum_I \Big(\frac{\partial V_{PS}}{\partial \bX_I}\Big)_{\bm{\Pi}^{\textit{eff}}}
    &=& -\left< \Psi \middle| \frac{\partial \hat{V}_{PS}}{\partial \bG}\middle|\Psi \right> = -\Big(\frac{\partial {V}_{PS}}{\partial \bG}\Big)_{{\bm{\Pi}}^{\textit{eff}}}  \label{eq:estdinv}
\end{eqnarray}
In the last equality above, we have used the Hellmann-Feynman equality (which follows because $\ket{\Psi}$ is an eigenstate of $\hat{V}_{PS}$). Therefore, on account of Eq. \ref{eq:eps_g_grad}, the final expression is zero:
\begin{eqnarray}
\sum_I \Big(\frac{\partial V_{PS}}{\partial \bX_I}\Big)_{\bm{\Pi}^{\textit{eff}}}   &=& 0 \label{eq:estdinv2}
\end{eqnarray}

\subsection{Rotational Invariance of $E_{PS}$}\label{rot_inv_eps}
Next, let us prove the rotational invariance of $E_{PS}$ around the magnetic origin, i.e. $\Delta_{rot}E_{PS} =0$ where
\begin{eqnarray}  \Delta_{rot}E_{PS} &=& \sum_{I\beta\gamma}\epsilon_{z\beta\gamma}\Big(P_{I\beta} \left(\frac{\partial E_{PS}}{\partial P_{I\gamma}}\right)_\bX+ (X_{I\beta}-G_\beta) \left(\frac{\partial E_{PS}}{\partial X_{I\gamma}}\right)_\bP\Big) \\
& =&  \Delta_{rot}E_{T_n} + \Delta_{rot}E_{\Gamma}+ \Delta_{rot}E_{e} 
\end{eqnarray}
Using Hellmann-Feynman theorem we can write each term as,
\begin{eqnarray} 
\Delta_{rot}E_{T_n} 
&=& \sum_{I\beta\gamma}\epsilon_{z\beta\gamma} \left( \left< \Psi \middle| P_{I\beta} \left(\frac{\partial {T}_{n}}{\partial P_{I\gamma}}\right)_\bX \middle| \Psi \right>+ \left< \Psi \middle|(X_{I\beta}-G_\beta) \left(\frac{\partial {T}_{n}}{\partial X_{I\gamma}}\right)_\bP\middle| \Psi \right> \right) \nonumber \\ \\
\Delta_{rot}E_{\Gamma} &=&\sum_{I\beta\gamma}\epsilon_{z\beta\gamma} \left( \left< \Psi \middle| P_{I\beta} \left(\frac{\partial \hat{H}_\Gamma}{\partial P_{I\gamma}}\right)_\bX \middle| \Psi \right>+ \left< \Psi \middle|(X_{I\beta}-G_\beta) \left(\frac{\partial \hat{H}_{\Gamma}}{\partial X_{I\gamma}}\right)_\bP\middle| \Psi \right> \right) \nonumber \\ \\ 
\Delta_{rot}E_e &=&\sum_{I\beta\gamma}\epsilon_{z\beta\gamma} \left( \left< \Psi \middle| P_{I\beta} \left(\frac{\partial \hat{H}_{e}}{\partial P_{I\gamma}}\right)_\bX \middle| \Psi \right>+ \left< \Psi \middle|(X_{I\beta}-G_\beta) \left(\frac{\partial \hat{H}_{e}}{\partial X_{I\gamma}}\right)_\bP\middle| \Psi \right> \right)  \nonumber \\
\end{eqnarray}
where ${T}_n$ is kinetic energy defined in Eq. \ref{eq:Tnuc} but in phase-space picture we replace the quantum nuclear momentum operator with the classical nuclear momentum, $\hat{H}_{\hat{\Gamma}}$ is defined in Eq. \ref{eq:Hgamma}, and $\hH_e$ is defined in Eq. \ref{eq:He}.

\subsubsection{$\Delta_{rot}E_{T_n}$}
The rotation of nuclear kinetic energy term is given by:
\begin{eqnarray}
\Delta_{rot}E_{T_n} & =& 
\sum_{I\beta\gamma}\epsilon_{z\beta\gamma}\left< \Psi \middle| 2 P_{I\beta}  (P_{I\gamma}-q_I^{\textit{eff}}A(\bX_I)_\gamma)\middle| \Psi \right>\label{eq:rot_inv1}
\nonumber\\
&& + \sum_{I\beta\gamma}\epsilon_{z\beta\gamma}\left< \Psi \middle|2(X_{I\beta}-G_\beta) \sum_\alpha (P_{I\alpha}-q_I^{\textit{eff}}A(\bX_I)_\alpha)(-\frac{q_I^{\textit{eff}}}{2}\epsilon_{\alpha z\gamma}B_z)\middle| \Psi \right>\nonumber \\&& + \sum_{I\beta\gamma}\epsilon_{z\beta\gamma}\left< \Psi \middle|2(X_{I\beta}-G_\beta) \sum_{J \alpha} (P_{J\alpha}-q_J^{\textit{eff}}A(\bX_J)_\alpha)(-\frac{\partial q_J^{eff}}{\partial X_{I\gamma}})A(\bX_J)\middle| \Psi \right> \nonumber\\\label{eq:rot_inv2}
\end{eqnarray}
There are three terms in the above equation.
Using the definition of $\bA$ in Eq. \ref{eqn:vector_potential}, the first two terms cancel. The last term is zero by Eqs. \ref{eq:qeff_inv_rot} and \ref{eq:qeff_inv_trans}.

\subsubsection{$\Delta_{rot}E_{\Gamma}$} Next, we turn to the $\hat{\Gamma}$ terms. The sum of $\bX$ and $\bP$ gradients of the $\hat{H}_\Gamma$ term is given by: 
\begin{eqnarray}
\Delta_{rot}E_{{\Gamma}}  &=&-i\hbar \sum_{I\beta\gamma}\epsilon_{z\beta\gamma}\left< \Psi \middle| P_{I\beta}\hat{\Gamma}_{I\gamma}\middle| \Psi \right>-i\hbar\sum_{I\beta\gamma}\epsilon_{z\beta\gamma}\left< \Psi \middle|(X_{I\beta}-G_\beta)\sum_\alpha   \frac{\partial\Pi_{I\alpha}^{\textit{eff}}}{\partial X_{I\gamma}}\hat{\Gamma}_{I\alpha}\middle| \Psi \right> \nonumber
\\ && -i\hbar \sum_{I\beta\gamma}\epsilon_{z\beta\gamma}\left< \Psi \middle|(X_{I\beta}-G_\beta)\sum_\alpha \Pi_{I\alpha}^{\textit{eff}} \frac{\partial \hat{\Gamma}_{I\alpha}}{\partial X_{I\gamma}}\middle| \Psi \right> \nonumber\\&&-\hbar^2 \sum_{I\beta\gamma}\epsilon_{z\beta\gamma}\left< \Psi \middle|2(X_{I\beta}-G_\beta)\sum_\alpha   \frac{\partial\hat{\Gamma}_{I\alpha}}{\partial X_{I\gamma}}\hat{\Gamma}_{I\alpha}\middle| \Psi \right> \label{eq:rot_inv3}
\end{eqnarray}
There are four terms above. For the second term, involving the gradient of $\bm{\Pi}^{\textit{eff}}$,  we use the fact that the total effective charge is constant when the whole molecule is rotated or translated (Eqs. \ref{eq:qeff_inv_rot} and \ref{eq:qeff_inv_trans}). Further, if we use the  rotational property of $\bm{\hat{\Gamma}}$ specified in Eq. \ref{eq:Gamma_uv4_nb}, the result is:  
\begin{eqnarray}
\Delta_{rot}E_{{\Gamma}} 
&=& -i\hbar \sum_{I\beta\gamma}\epsilon_{z\beta\gamma}\left< \Psi \middle| P_{I\beta}\hat{\Gamma}_{I\gamma}\middle| \Psi \right>  \nonumber\\&&-i\hbar\sum_{I\beta\gamma}\epsilon_{z\beta\gamma}\left< \Psi \middle|(X_{I\beta}-G_\beta)\sum_\alpha (-\frac{q_I^{\textit{eff}}}{2}\epsilon_{\alpha z\gamma}B_z)\hat{\Gamma}_{I\alpha}\middle| \Psi \right> \nonumber \nonumber\\
&&-\sum_{I\alpha}\Pi_{I\alpha}^{\textit{eff}}\left< \Psi \middle| \left[\hat{l}^z,  \hat{\Gamma}_{I\alpha}\right] \middle| \Psi \right>+i\hbar \sum_{I\delta\alpha}\epsilon_{\delta z \alpha}\left< \Psi \middle| \Pi_{I\alpha}^{\textit{eff}}  \hat{\Gamma}_{I\delta}\middle| \Psi \right> \nonumber\\&&- \hbar^2 \sum_{I\alpha}\left< \Psi \middle|2\left[\frac{\hat{l}^z}{i\hbar},\hat{\Gamma}_{I\delta}\right]\hat{\Gamma}_{I\alpha}\middle| \Psi \right>+\hbar^2 \sum_{I\alpha\delta}\epsilon_{\delta z\alpha}\left< \Psi \middle| \hat{\Gamma}_{I\delta} \hat{\Gamma}_{I\alpha}\middle|\Psi \right> \label{eq:rot_inv5}
\end{eqnarray}
The first, second, and fourth terms cancel in Eq. \ref{eq:rot_inv5}, and the sixth term is zero by symmetry. If we combine the third and fifth terms, the 
 can be written as a commutator of electronic angular momentum and the  $\hat{H}_{\Gamma}$ operator: 
\begin{eqnarray}
\Delta_{rot}E_{{\Gamma}} &=& \left< \Psi \middle| \left[\frac{\hat{l}^z}{i\hbar},  \hat{H}_{\Gamma}\right] \middle| \Psi \right>
\end{eqnarray}

\subsubsection{$\Delta_{rot}E_{e}$}
Finally, the last term is the gradient of the electronic energy $E_e$ to rotations. Noting that $\hH_e$ does not depend on $\bP$ and that the total angular momentum commutes with the electronic Hamiltonian (this is the rotational analogue of Eq. \ref{eq:Gamma_uv4_nb} above),
\begin{eqnarray}
    \Big[-i\hbar\sum_{I}\left((\bm{X}_I-\bG) \times\pp{}{\bm{X}_I}\right)_{\gamma} + ((\br-\bG)\times \bp)_{\gamma}, \hat{H}_e\Big] = 0,
\end{eqnarray}
we can simplify this term as:
\begin{eqnarray}
\Delta_{rot}E_{e}
&=& \sum_{I\beta\gamma}\epsilon_{z\beta\gamma} \left< \Psi \middle|(X_{I\beta}-G_\beta) \left(\frac{\partial \hat{H}_{e}}{\partial X_{I\gamma}}\right)_\bP \middle| \Psi \right>\\
&=& \left< \Psi \middle|\left[\frac{\hat{l}_z}{i\hbar},\hat{H}_{e}\right]\middle| \Psi \right>
\end{eqnarray}

\subsubsection{Summing Up All of the Terms}

Adding all the non-zero terms together yields:
\begin{eqnarray}
    \Delta_{rot}E_{{\Gamma}} + \Delta_{rot}E_{e} &=& \left< \Psi \middle| \left[\frac{\hat{l}_z}{i\hbar},  \hat{H}_{\Gamma}\right] \middle| \Psi \right> + \left< \Psi |\left[\frac{\hat{l}_z}{i\hbar},\hat{H}_{e}\right]\middle| \Psi \right>\\ 
    &=& \left< \Psi \middle| \left[\frac{\hat{l}_z}{i\hbar},  \hat{V}_{PS}\right] \middle| \Psi \right>
\end{eqnarray}
Finally, noting that $\ket{\Psi}$ is an eigenstate of $\hat{H}_{PS}$, we conclude that:
\begin{eqnarray}
\Delta_{rot} E_{PS} &=&\sum_{I\alpha}\frac{1}{i\hbar}\Big(\left< \Psi \middle|\hat{l}_z \hat{V}_{PS}\middle| \Psi \right>-\left< \Psi \middle|\hat{V}_{PS} \hat{l}_z \middle| \Psi \right> \Big) = 0 \label{eq:EPSrot}
\end{eqnarray}
Hence, we have shown that the phase-space electronic energy is rotationally invariant.


\subsection{Pseudomomentum Conservation}\label{pm_conserve}
The relations above enable  us to prove conservation of  the pseudomomentum.
For a collection of particles (with charges $q^{\textit{eff}}(\bm{X})$) in a magnetic field, the total conserved quantity should be the sum of the pseudomomentum of the nuclei and the pseudomomentum like term for electrons. We will define these quantities as follows:
\begin{eqnarray}
  K^\alpha_{mol} &=&
  K^\alpha_{n} +
  k^{\alpha}_{e} \label{eq:joe1}
  \\
  K^\alpha_{n} &=&
  \sum_I {P_{I\alpha}}+ \frac{1}{2}\sum_I q^{\textit{eff}}_I(\bX)(\bB\times (\bX_I-\bG))_{\alpha}-i\hbar \sum_{I} \left< \Psi \middle| \hat{\Gamma}_{I\alpha} \middle| \Psi \right> \label{eq:joe2}\\ 
  k^{\alpha}_{e} &=&  \sum_{I}\left< \Psi \middle| \hat{\Theta}_I\hat{k}_I^{\alpha} \middle| \Psi \right> \label{eq:joe3}
\end{eqnarray}
Plugging in these expressions, the result is:
\begin{eqnarray}
   \label{eq:pseudomom2} K^\alpha_{mol} &=& \sum_I {P_{I\alpha}}+ \frac{1}{2}\sum_I q^{\textit{eff}}_I(\bX)(\bB\times (\bX_I-\bG))_{\alpha}-i\hbar \sum_{I} \left< \Psi \middle| \hat{\Gamma}_{I\alpha} \middle| \Psi \right> \nonumber \\&&+ \sum_{I} \left< \Psi \middle| \hat{\Theta}_I\hat{k}_I^{\alpha} \middle| \Psi \right> \\ &=& \sum_I {P_{I\alpha}}+ \frac{1}{2}\sum_I q^{\textit{eff}}_I(\bX)(\bB\times (\bX_I-\bG))_{\alpha}\label{eq:kmol}
\end{eqnarray}
Note that the last two terms in Eq. \ref{eq:pseudomom2} cancel using condition 1 in Eq. \ref{eq:Gamma_uv1}.

Now, the form of Eq. \ref{eq:joe3} might look unnerving (insofar as we use the relative electronic pseudomomentum rather than the absolute pseudomomentum). Nevertheless, it is crucial to note that here we have used {\em effective} (not raw) charges in Eq. \ref{eq:joe2}. As we will show below, in Sec. \ref{justify}, the pseudomomentum in Eq. \ref{eq:joe1} is actually quite close to the raw quantity one would like to conserve. 

At this point, in order to prove that the pseudomomentum is a constant of motion, we can evaluate the time derivative of Eq. \ref{eq:kmol}, using the equation of motions in Eq. \ref{eq:Estdgrad} (noting that, of course, the gauge origin $\bG$ is fixed and its time derivative is zero):
\begin{eqnarray}
   \frac{d  K^\alpha_{mol}}{dt} &=& \sum_I \dot P_{I\alpha} + \frac{1}{2}\sum_Iq_I^{\textit{eff}}({\bX})(\bB \times \dot \bX_I)_{\alpha} \\ 
   &=&-\sum_{IJ\beta} \frac{\Pi^{\textit{eff}}_{J\beta}}{M_J}\Big(- \frac{1}{2} \frac{\partial q^{\textit{eff}}_J(\bX)}{\partial X_{I\alpha}}(\bB\times \bX_J)_{\beta}- \frac{1}{2} q^{\textit{eff}}_J(\bX)\frac{\partial (\bB\times \bX_J)_{\beta}}{\partial X_{I\alpha}}\Big)\nonumber\\
   &&+i\hbar \sum_{IJ\beta} \left< \Psi \middle|  \frac{1}{M_J}\Bigg(- \frac{1}{2} \frac{\partial q^{\textit{eff}}_J(\bX)}{\partial X_{I\alpha}} (\bB\times \bX_J)_{\beta} - \frac{1}{2} q^{\textit{eff}}_J(\bX)\frac{\partial (\bB\times \bX_J)_{\beta} }{\partial X_{I\alpha}}\Bigg) \hat{\Gamma}_{J\beta}\middle| \Psi \right>\nonumber\\&&-\sum_I\left(\frac{\partial V_{PS}}{\partial X_{I \alpha}}\right)_{\Pi^\textit{eff}}+ \frac{1}{2}\sum_Iq_I^{\textit{eff}}({\bX})(\bB \times \dot \bX_I)_{\alpha}
\end{eqnarray}
At this point, we invoke Eq. \ref{eq:qeff_inv_trans} (which states that $q_I^{eff}$ is translationally invariant)
 and Eq. \ref{eq:estdinv2} (which states that $V_{PS}$ is translationally invariant) so that two terms vanish, and we find:
\begin{eqnarray}
\frac{d  K^\alpha_{mol}}{dt} &=&-\sum_{IJ\beta} \frac{\Pi^{\textit{eff}}_{J\beta}}{M_J}\Big(- \frac{1}{2} q^{\textit{eff}}_J(\bX)\frac{\partial (\bB\times \bX_J)_{\beta}}{\partial X_{I\alpha}}\Big)_P \nonumber\\
   &&+i\hbar \sum_{IJ\beta} \left< \Psi \middle|  \frac{1}{M_J}\Big(- \frac{1}{2} q^{\textit{eff}}_J(\bX)\frac{\partial (\bB\times \bX_J)_{\beta} }{\partial X_{I\alpha}}\Big)_P \hat{\Gamma}_{J\beta}\middle| \Psi \right>\nonumber\\&&+ \frac{1}{2}\sum_Iq_I^{\textit{eff}}({\bX})(\bB \times \dot \bX_I)_{\alpha}
\end{eqnarray}
Next, using Eq. \ref{eq:xdot},  we expand $\dot{\bX}_{I}$ 
in the equation above to simplify:
\begin{eqnarray}
   \frac{d K^\alpha_{mol}}{dt} &=&\frac{1}{2}\sum_{IJ\beta}  q^{\textit{eff}}_J(\bX) \epsilon_{\beta z \alpha}B_z\frac{\Pi^{\textit{eff}}_{J\beta}}{M_J}\delta_{IJ}-\frac{i\hbar}{2}\sum_{I J\beta}q^{\textit{eff}}_J(\bX)\epsilon_{\beta z \alpha}B_z \left< \Psi \middle| \frac{ \hat{\Gamma}_{J\beta}}{M_{J}}\middle| \Psi \right> \delta_{IJ}\nonumber\\
   &&+ \frac{1}{2}\sum_{I\beta}q^{\textit{eff}}_I(\bX) \epsilon_{\alpha z \beta}B_z\frac{\Pi^{\textit{eff}}_{I\beta}}{M_{I}} - \frac{i\hbar}{2}\sum_{I\beta} q^{\textit{eff}}_I(\bX) \epsilon_{\alpha z \beta}B_z \left< \Psi \middle|\frac{ \hat{\Gamma}_{I\beta}}{M_{I}} \middle| \Psi \right>\\
   &=& 0\label{eq:pseudoconsv}
\end{eqnarray}
Therefore, the pseudomomentum in the arbitrary direction $\alpha$ is conserved in a uniform magnetic field. Note that, for this proof, all that we have required is translational invariance of the phase-space energy and effective charges.

\subsection{Angular momentum conservation}\label{ang_conserve}
An analogous argument can be used to prove the conservation of the canonical angular momentum.
We posit that the conserved quantity should be the sum of the canonical angular momentum of the nuclei (albeit with a screened effective charge)  plus a pseudoangular momentum like term for the electronic cloud around each charge center. Mathematically, we write:
\begin{eqnarray}
    L_{mol}^z &=& L_{n}^z + L_{e}^z \\
    L_{n}^z &=& \sum_I \Bigg((\bX_I-\bG) \times \Big( \bm{\Pi}^{\textit{eff}}_{I} - i\hbar \left< \Psi \middle|\bm{\hat{\Gamma}}_{I}\middle| \Psi \right>+\frac{q^{\textit{eff}}_I(\bX)}{2}(\bB\times (\bX_I-\bG))\Big)\Bigg)_z\nonumber\\\\
    L_{e}^z &=& \sum_I \left< \Psi \middle|( \hbr - \bG)\times \frac{1}{2} \left( \hat{\Theta}_I \hat{\bm{k}}_{Iz} +  \hat{\bm{k}}_{Iz} \hat{\Theta}_I \right) \middle| \Psi \right>
    \label{eq:funnyL}
\end{eqnarray}
so that
\begin{eqnarray}
L_{mol}^z
    &=& \sum_I \Bigg((\bX_I-\bG) \times \Big(M_I  \bm{\dot{X}}_I+\frac{q_I^{\textit{eff}}(\bX)}{2}(\bB\times (\bX_I-\bG))\Big)\Bigg)_z \nonumber\\
    & + & \sum_I \left< \Psi \middle|( \hbr - \bG)\times \frac{1}{2} \left( \hat{\Theta}_I \hat{\bm{k}}_{Iz} +  \hat{\bm{k}}_{Iz} \hat{\Theta}_I \right) \middle| \Psi \right> \label{eq:Ltot}
\end{eqnarray}

As for the pseudomomentum, we will prove in Sec. \ref{justify} that, despite the awkward inclusion of $\hat{\bm{k}}_I$ and the the use of the screened charges $q^{\textit{eff}}$, the total quantity delineated above does in fact approximate the usual form for the canonical angular momentum. 
In the meantime, we notice that this expression can be simplified by plugging in the definition of $\dot{\bX}_I$ (in Eq. \ref{eq:xdot}):
\begin{eqnarray}
 L^{z}_{mol} &=&  \sum_{I\beta\gamma} \epsilon_{z\beta\gamma}(X_{I\beta}-G_{\beta})M_{I} \dot{X}_{I\gamma} + \sum_{I\beta\gamma} \epsilon_{z\beta\gamma}(X_{I\beta}-G_{\beta})\frac{q_I^{\textit{eff}}(\bX)}{2}(\bB\times (\bX_I-\bG))_\gamma \nonumber\\&&+  \sum_I\left< \Psi \middle|{(\hat{\Theta}_I( \hbr- \bG)\times \hat{\bm{k}}_I)}_{z} \middle| \Psi \right>  \\
&=& \sum_{I\beta\gamma} \epsilon_{z\beta\gamma}(X_{I\beta}-G_{\beta})M_{I}\Big( \frac{ \Pi^{\textit{eff}}_{I\gamma}}{M_{I}} - i\hbar\left< \Psi \middle|\frac{ \hat{\Gamma}_{I\gamma}}{M_{I}} \middle| \Psi \right>\Big) \nonumber\\
&&+ \sum_{I\beta\gamma} \epsilon_{z\beta\gamma}(X_{I\beta}-G_{\beta})\frac{q_I^{\textit{eff}}(\bX)}{2}(\bB\times (\bX_I-\bG))_\gamma \nonumber\\
&&+ \sum_I\left< \Psi \middle|  \Big(( \hbr- \bG)\times \frac{1}{2} ( \hat{\Theta}_I \hat{\bm{k}}_{I} +  \hat{\bm{k}}_{I} \hat{\Theta}_I ) \Big)_z\middle| \Psi \right> 
 \label{eq:lst} 
\end{eqnarray}
Using Eq. \ref{eq:Gamma_uv3} to cancel the term involving $\bGamma$, we find:
 \begin{eqnarray}
 L^{z}_{mol}
&=& \sum_{I\beta\gamma} \epsilon_{z\beta\gamma}(X_{I\beta}-G_{\beta}) \Pi^{\textit{eff}}_{I\gamma}
\nonumber\\&&+ \sum_{I\beta\gamma} \epsilon_{z\beta\gamma}(X_{I\beta}-G_{\beta})\frac{q_I^{\textit{eff}}(\bX)}{2}(\bB\times (\bX_I-\bG))_\gamma 
 \label{eq:lst2} 
\end{eqnarray}
and further expanding $\Pi_{I\gamma}^{\textit{eff}}$ (using Eq. \ref{eq:pieffdefn}) yields:
\begin{eqnarray}
L^{z}_{mol} 
&=& \sum_{I\beta\gamma} \epsilon_{z\beta\gamma}(X_{I\beta}-G_{\beta})(P_{I\gamma}-\frac{q_I^{\textit{eff}}(\bX)}{2}(\bB\times (\bX_I-\bG))_\gamma)\nonumber\\&&+ \sum_{I\beta\gamma} \epsilon_{z\beta\gamma}(X_{I\beta}-G_{\beta})\frac{q_I^{\textit{eff}}(\bX)}{2}(\bB\times (\bX_I-\bG))_\gamma \\
&= & \sum_{I\beta\gamma} \epsilon_{z\beta\gamma}(X_{I\beta}-G_{\beta})P_{I\gamma}
\end{eqnarray}

Therefore,
\begin{eqnarray}
\frac{d}{dt} L^{z}_{mol}&=& \frac{d}{dt} \Big[ \sum_{I\beta\gamma} \epsilon_{z\beta\gamma}(X_{I\beta}-G_{\beta})P_{I\gamma}\Big]\\
&=&  \sum_{I\beta\gamma}\epsilon_{z\beta\gamma}\Big(\dot X_{I\beta} P_{I\gamma} + (X_{I\beta}-G_{\beta})\dot P_{I\gamma}\Big) 
\\
 &=&  -\sum_{I\beta\gamma}\epsilon_{z\beta\gamma}\Big(P_{I\beta} \left(\frac{\partial E_{PS}}{\partial P_{I\gamma}}\right)_X+ (X_{I\beta}-G_\beta) \left(\frac{\partial E_{PS}}{\partial X_{I\gamma}}\right)_P\Big)\label{eq:rot_inv}
\end{eqnarray}

Note that,  in order for angular momentum to be conserved, the phase-space energy needs only to be rotationally invariant around the $z-$axis -- which we have already proven in Sec. \ref{rot_inv_eps}. 

\section{A Basis Free Ansatz for ETFs and ERFs}\label{ETF-ERF}

It remains only to specify our proposed form for the $\bGamma$ operators -- so as to both approximate the derivative couplings and satisfy the relevant constraints in Eqs. \ref{eq:Gamma_uv1}-\ref{eq:Gamma_uv3} and Eqs. \ref{eq:Gamma_uv2_nb} and \ref{eq:Gamma_uv4_nb} above. 
As described in Refs. \citenum{coraline_pssh,tao2024basis} (in the case without a magnetic field), the natural candidates are electron translation factors (ETFs)\cite{fatehi_etf,schneiderman:1969:pr:etf} and electron rotation factors (ERFs)\cite{tao2024basis,tian_erf,athavale:2023:erf}.
In general, we will write 
\begin{eqnarray}
    \hat{\bGamma} = \hat{\bGamma}' + \hat{\bGamma}'' 
    \label{eq:gammatotal}
\end{eqnarray}
While often conceived of as a means to improve rescaling direction in surface hopping calculations \cite{wu2024linear, bian2024surfacehopping},  ETFs and ERFs more generally specify how an electron is dragged along with the nuclei on which the atomic orbitals are attached. While ETFs can be constructed in strictly local fashion\cite{fatehi_etf}, the ERFs are defined in a semi-local fashion, since they depend on the angular momentum\cite{tian_erf}. Using these modified derivative couplings leads to conservation of linear and angular momentum during an adiabatic propagation. 


\subsection{Modified ETF expression}

In order to satisfy Eq. \ref{eq:Gamma_uv1}, our first step is to define the translation factors $\bGamma'$. To that end, we  will define:
\begin{eqnarray}
     \bm{\hat{\Gamma}}_I' &=& \frac{1}{2i\hbar}\left( \hat{\Theta}_I{\hat{\bm{ k}}_I}+{\hat{\bm{ k}}_I}\hat{\Theta}_I\right)\label{eq:etf_trans}\\
 &=& \frac{1}{2i\hbar}\left( \hat{\Theta}_I{\hat{\bm{\pi}}}+{\hat{\bm{\pi}}}\hat{\Theta}_I\right)-\frac{1}{i\hbar}\hat{\Theta}_I(e\bm{B}\times (\bm{\hat{r}}-\bm{X}_I))\label{eq:etf}
\end{eqnarray}
This ETF term is translationally, rotationally, and gauge invariant. 

\subsection{Modified ERF expression}
Next, Eqs. \ref{eq:Gamma_uv1} and \ref{eq:Gamma_uv3} above are equivalent to  the following two conditions:
\begin{align}
   \sum_I\bm{ \hat{\Gamma}}_I &= \sum_I \left(\bm{ \hat{\Gamma}}'_I+\bm{\hat{\Gamma}}''_I\right) = \sum_I\frac{1}{2i\hbar}(\hat{\Theta}_I\hat{\bm{k}}_I + \hat{\bm{k}}_I \hat{\Theta}_I)\label{eq:etf_munu}
\end{align}
and
\begin{align}
    \sum_I (\bm{X}_I -\bm{G})\times\bm{ \hat{\Gamma}}_I = \sum_I (\bm{X}_I -\bm{G})\times\left(\bm{ \hat{\Gamma}}'_I+\bm{ \hat{\Gamma}}''_I\right) = \sum_I\frac{\hat{\Theta}_I((\bm{\hat{r}}-\bm{G})\times \hat{\bm{k}}_I)}{i\hbar} \label{eq:erf_munu}
\end{align}
Given that $\bm{ \hat{\Gamma}}'$ already satisfies the relation in Eq. \ref{eq:etf_munu}, it follows that the $\bm{ \hat{\Gamma}}''$ term must satisfy:
\begin{eqnarray}
    \sum_I\bm{\hat{\Gamma}}''_I &=& 0 \label{eq:sum_gamma}
\end{eqnarray}
For the constraint in Eq. \ref{eq:erf_munu}, we note that $\bGamma''$ must further satisfy:  
\begin{eqnarray}
\sum_{I}(\bm X_{I}-\bm{G}) \times \mathbf{\hat{\Gamma}}''_{I} &=& \sum_I\frac{\hat{\Theta}_I}{i\hbar}(\hbm{r}-\bm{G})\times\hat{\bm{k}}_I - \sum_I (\bm{X}_I-\bm{G})\times\bm{\hat{\Gamma}}_I'\label{eq:angmom_Gamma'}
\end{eqnarray}
Substituting in the $\bm{\hat{\Gamma}}'_I$ expression from Eq. \ref{eq:etf_trans}, Eq. \ref{eq:angmom_Gamma'} becomes:  
\begin{eqnarray}
    \sum_I (\bm{X}_I-\bm G)\times\bm{\hat{\Gamma}}_I''
    &=& \sum_I\frac{\hat{\Theta}_I}{i\hbar}(\hbm{r}-\bm G)\times\hat{\bm{k}}_{I} \nonumber\\&&- \sum_I (\bm{X}_I-\bm G)\times(\frac{1}{2i\hbar}(\hat{\Theta}_I(\hbr) \hat{\bm{k}}_{I} + \hat{\bm{k}}_{I}\hat{\Theta}_I(\hbr)))\\ &=& \sum_I(\hbm{r}-\bm G)\times(\frac{1}{2i\hbar}(\hat{\Theta}_I(\hbr) \hat{\bm{k}}_{I} + \hat{\bm{k}}_{I}\hat{\Theta}_I(\hbr)))\nonumber \\&& - \sum_I (\bm{X}_I-\bm G)\times(\frac{1}{2i\hbar}(\hat{\Theta}_I(\hbr) \hat{\bm{k}}_{I} + \hat{\bm{k}}_{I}\hat{\Theta}_I(\hbr)))\\&=& \frac{1}{2i\hbar}\sum_I(\hbr-\bm X_I)\times(\hat{\Theta}_I(\hbr) \hat{\bm{k}}_{I} + \hat{\bm{k}}_{I}\hat{\Theta}_I(\hbr)) \label{eq:erf_giao_temp}
\end{eqnarray}
Using the same approach as in Ref. \citenum{tian_erf} for the ERF term, we will now guess that a meaningful $\hat{\bGamma}''$ term is:
\begin{align}
  \hat{\bm{\Gamma}}_I^{''} = \sum_{J}  \zeta_{IJ}\left(\bm{X}_I -\bm{X}^0_{J}\right)\times \left(\bm{K}_J^{-1}\frac{1}{2i\hbar}(\hbm{r}-\bm X_J)\times(\hat{\Theta}_J(\hbr) \hat{\bm{k}}_{J} + \hat{\bm{k}}_{J}\hat{\Theta}_J(\hbr))\right) \label{eq:final_erf}
\end{align}
where the $  \zeta_{IJ}$, $\bm{X}_J^0$, and $\bm{K}_J$ are defined as follows:
\begin{eqnarray}
      \zeta_{IJ} &=& e^{-|\bm{X}_I-\bm{X}_J|^2/2(\sigma_I+\sigma_J)^2}\label{eq:zeta}\\
    \bm{X}_{J}^0 &=&\frac{\sum_I  \zeta_{IJ}\bm{X}_I}{\sum_I  \zeta_{IJ}}\\
    \bm{K}_J &=&\sum_I  \zeta_{IJ}\left(\bm{X}_I\bm{X}_I^\top-\bm{X}_J^0\bm{X}_J^{0\top}-(\bm{X}_I^\top\bm{X}_I-\bm{X}_J^{0\top}\bm{X}_J^0)\mathcal{I}_3\right)
\end{eqnarray}
Clearly, the definition of $\bm{ \hat{\Gamma}}''_{I}$ in Eq. \ref{eq:final_erf} already satisfies the constraint given in Eq. \ref{eq:sum_gamma} because:
\begin{align}
\sum_I\bm{ \hat{\Gamma}}_I'' = \sum_{IJ}  \zeta_{IJ}\left(\bm{X}_I -\bm{X}^0_{J}\right)\times \left(\bm{K}_J^{-1}\hat{\bm{ J}}_J\right) = 0
\end{align}
for any function $\hat{\bm{J}}_J$. Furthermore, an elementary calculation reveals that the definition in Eq. \ref{eq:final_erf} also satisfies Eq. \ref{eq:erf_giao_temp}.
Finally and most importantly, note that Eq. \ref{eq:final_erf}  is also gauge independent, as well as translationally and rotationally invariant.

Lastly, although we have not addressed spin here, given that $(i)$ in the absence of a magnetic field, the total angular momentum (including spin) is conserved and $(ii)$ we wish the present phase-space approach to reduce to the expression from Ref. \citenum{tao2024basis} when $\bB = 0$, we posit that the most general form of $\hat{\bGamma}''$ must be:
\begin{align}
  \hat{\bm{\Gamma}}_I^{''} = \sum_{J}  \zeta_{IJ}\left(\bm{X}_I -\bm{X}^0_{J}\right)\times \left(\bm{K}_J^{-1}\frac{1}{2i\hbar} \left[ (\hbm{r}-\bm X_J)\times(\hat{\Theta}_J(\hbr) \hat{\bm{k}}_{J} + \hat{\bm{k}}_{J}\hat{\Theta}_J(\hbr)) + 2 \hat{\bm{s}} \right]\right) \label{eq:final_erf_s}
\end{align}
where $\hat{\bm{s}}$ is the electronic spin operator. Future work will be essential to incorporate spin-orbit coupling and spin-Zeeman into the present formalism.\cite{neese2005efficient,tang2024exact}

\subsection{Form of Effective Charge $\bm{q}^{\textit{eff}}$}\label{qeff_define}
At this point, let us define the screened charge $q_I^{\textit{eff}}$  in Eq. \ref{eq:PSH}.  To that end, we will fix
 $q^{\textit{eff}}_I$ as:
\begin{eqnarray}
    q^{\textit{eff}}_I(\bm{X}) &=&   Q_Ie -e\left< \Psi_0 \middle|\hat{\Theta}_I(\hbr,\bm{X})\middle| \Psi_0 \right>\label{eq:qeffdefn}
\end{eqnarray}
Here, the phase-space wavefunction $\ket{\Psi_0}$ is defined at zero magnetic field. Note that $q_I^{\textit{eff}}$ clearly satisfies Eqs. \ref{eq:qeff_inv_trans} 
and \ref{eq:qeff_inv_rot} above.

\subsection{Reinterpreting the Form of the Conserved Pseudomomentum and Angular Momentum}\label{justify}

In Secs. \ref{pm_conserve} and \ref{ang_conserve} above, we posited forms for the pseudomomentum and canonical angular momentum that did not necessarily appear physically obvious (though these forms did make clear that they obeyed translational and rotational invariance). Here, we wish to rewrite those  same expressions in a different fashion so as to make more obvious their physical meaning.
The total pseudomomentum according to Eq. \ref{eq:pseudomom2} is given by:
\begin{eqnarray}
    K^\alpha_{mol} &=& \sum_I {P_{I\alpha}}+ \frac{1}{2}\sum_I q^{\textit{eff}}_I(\bX)(\bB\times (\bX_I-\bG))_{\alpha}-i\hbar \sum_{I} \left< \Psi \middle| \hat{\Gamma}_{I\alpha} \middle| \Psi \right> \nonumber \\&&+ \sum_{I} \left< \Psi \middle| \hat{\Theta}_I{\hat{k}}_I^{\alpha} \middle| \Psi \right> \\
    &=& \sum_I {P_{I\alpha}}+ \frac{1}{2}\sum_I q^{\textit{eff}}_I(\bX)(\bB\times (\bX_I-\bG))_{\alpha}-i\hbar \sum_{I} \left< \Psi \middle| \hat{\Gamma}_{I\alpha} \middle| \Psi \right> \nonumber \\&&+  \left< \Psi \middle| \hat{k}^{\alpha}\middle| \Psi \right> + \sum_{I} e  \left< \Psi \middle|\hat{\Theta}_I(\hbr,\bm{X})\middle| \Psi \right>(\bB \times (\bX_I-\bG))_\alpha 
\end{eqnarray}
At this point, if we ignore the distinction between $\ket{\Psi}$ and $\ket{\Psi_0}$,  
\begin{eqnarray}
\label{eq:approx_joe}
\ket{\Psi} \approx \ket{\Psi_0}  
\end{eqnarray}
we find that:
\begin{eqnarray}
        K^\alpha_{mol}& = & \sum_I \left( \Pi^{\textit{kin}}_{I \alpha} + Q_Ie(\bB\times (\bX_I-\bG))_{\alpha} \right) + \left< \Psi \middle| \hat{k}^\alpha\middle| \Psi \right>
\end{eqnarray}
where $\bm{\Pi}^{\textit{kin}}_I$ is defined in Eq. \ref{eq:pikindefn}. Therefore, at the level of  the approximation in Eq. \ref{eq:approx_joe}, the total pseudomomentum that is conserved in the phase-space picture is indeed the effective total pseudomomentum of the nuclei and electrons.

Likewise, we have postulated that the total canonical angular momentum (Eq. \ref{eq:Ltot}) is given by:
\begin{eqnarray}
    L_{mol}^z &=& \sum_I \Bigg((\bX_I-\bG) \times \Big( \bm{\Pi}^{\textit{eff}}_{I} - i\hbar \left< \Psi \middle|\bm{\hat{\Gamma}}_{I}\middle| \Psi \right>+\frac{q_I^{\textit{eff}}(\bX)}{2}(\bB\times (\bX_I-\bG))\Big)\Bigg)_z \nonumber\\&&+ \sum_I \left< \Psi \middle| \hat{\Theta}_I{(( \hbr- \bG)\times \hat{\bm{k}}_I)}_{z} \middle| \Psi \right> \\
    &=& \sum_I \Bigg((\bX_I-\bG) \times \Big( \bm{\Pi}^{\textit{kin}}_{I} +\frac{Q_Ie}{2}(\bB\times (\bX_I-\bG))\Big)\Bigg)_z \nonumber \\
    & & -  \frac{e}{2} \left< \Psi \middle|\hat{\Theta}_I\middle| \Psi \right>
    \Big((\bX_I-\bG) \times (  \bB\times (\bX_I-\bG))\Big)_z
    \nonumber\\&&+ \sum_I \left< \Psi \middle|\hat{\Theta}_I{\Big(( \hbr- \bG)\times (\hat{\bm{\pi}} - e\bB \times (\hbr-\bX_I))\Big)}_{z} \middle| \Psi \right>
\end{eqnarray}
 If we again make the approximation in Eq. \ref{eq:approx_joe} as well as further assert that for all linear and quadratic functions $f(\hbr)$, $\left< \Psi \middle|f(\hbr-\bX_I) \hat{\Theta}_I \middle| \Psi \right> \approx 0$ (since $\left< \Psi \middle|\hbr \hat{\Theta}_I \middle| \Psi \right> \approx \bX_I$),
 it follows that:
 \begin{eqnarray}
    L_{mol}^z 
    &\approx& \sum_I \Bigg((\bX_I-\bG) \times \Big( \bm{\Pi}^{\textit{kin}}_{I} +\frac{Q_Ie}{2}(\bB\times (\bX_I-\bG))\Big)\Bigg)_z \nonumber \\
    & &+ \sum_I \left< \Psi \middle|\hat{\Theta}_I \Big(( \hbr- \bG)\times (\hat{\bm{\pi}} -\frac{e}{2} (\bB\times (\bX_I-\bG)))\Big)_{z} \middle| \Psi \right> \\
     & \approx & 
     \sum_I \Bigg((\bX_I-\bG) \times \Big( \bm{\Pi}^{\textit{kin}}_{I} +\frac{Q_Ie}{2}(\bB\times (\bX_I-\bG))\Big)\Bigg)_z \nonumber \\
    & &+  \left< \Psi \middle| \Big(( \hbr- \bG)\times (\hat{\bm{\pi}} -\frac{e}{2} (\bB\times (\hbr-\bG)))\Big)_{z} \middle| \Psi \right> 
\end{eqnarray}
In other words, up to a reasonable approximation, the total conserved quantity does mimic the sum of the effective canonical angular momenta of the nuclei and  electrons in the direction of the magnetic field. 

\section{ Results For The Hydrogen Atom Hamiltonian}\label{results}

Eqs. \ref{eq:PSH},  \ref{eq:gammatotal}, \ref{eq:etf} and \ref{eq:final_erf} are the final set of equations that encapsulate our proposed phase-space electronic Hamiltonian.
That being said, 
there are many possible phase-space electronic Hamiltonians that conserve momentum (and, in this case, total pseudomomentum and total canonical angular momentum in direction of the $\bB$ field) with gauge invariance\cite{wu2024linear}. After all, both of these features more or less arise from translational and rotational invariance. 
Now, as described in Sec. \ref{semi_class_ham_PS}, we  fashioned our phase-space approach and $\hat \bGamma$ operator as an attempt to approximate the derivative coupling (so that the phase-space eigenstates represent an improvement and/or correction to BO eigenstates). Nevertheless, one can always wonder how good an approximation we have made and/or question the value of the proposed phase space electronic Hamiltonian. Thus, in short, despite the appealing formal properties offered above, one must ask: is there an easy means (without comparing to experiment) to further test the practical value and accuracy of the proposed phase-space electronic Hamiltonian?  
 
 From our perspective, the strongest means to test any phase-space electronic Hamiltonian is to demand that the eigenspectrum must be exact for the simplest case, a hydrogen atom. The hydrogen atom  is very relevant for applications in astrophysics, atomic spectroscopy and solid state physics \cite{Hexactsoln}. As is well known, there are no exact formulae for the eigenspectrum in a magnetic field (even though perturbative results are possible either in the limit of weak or strong magnetic fields (even though we can use perturbative results in the limit of weak magnetic field and adiabatic approximation in the limit of strong magnetic fields)\cite{Hexactsoln}), but one can reduce the Hamiltonian to a manageable analytic form all the same.
With this in mind, we will now show that the Hamiltonian in Eq. \ref{eq:PSH} (with $q^{\textit{eff}} = 0$) is indeed exact for the hydrogen atom.


\subsection{Standard (Exact) Hamiltonian}

For a hydrogen atom in a magnetic field,  the Hamiltonian is the sum of the kinetic energy of the electron, the kinetic energy of proton and the potential energy due to the interaction between the electron and proton ( here written in Cartesian coordinates) : 
\begin{eqnarray}
    \hat{H}_{exact} &=& \frac{1}{2M_n}(\hbP_n-\frac{e}{2}\bB\times \hbR_n)^2 + \frac{1}{2m_e}(\hbp_e+ \frac{e}{2}\bB\times \hbr_e)^2 + \hat{V}(\hbr_e-\hbR_n) 
\end{eqnarray}
If we separate the center of mass motion of the atom $(\bR,\bP_{R})$ from its internal motion $(\br,\bp_{r})$, the result is \cite{schmelcher1992regularity}:
\begin{eqnarray}
    \hat{H}_{exact} &=& \frac{1}{2M}(\hbP_R+\frac{e}{2}\bB\times \hbr)^2 + \frac{1}{2\mu}(\hbp_r+\frac{e}{2}\bB\times \hbR+ \frac{\mu}{\mu'}\frac{e}{2}\bB\times \hbr)^2+\hat{V}(\hbr)
\end{eqnarray}
where $\mu' = \frac{m_eM_n}{M_n-m_e}$ and the total mass is $M=M_n+m_e$.
Next, we write the above Hamiltonian in terms of the pseudomomentum (which in center of mass coordinates is expressed as $\hat{\bm{K}} = \hbP_R -\frac{e}{2}\bB\times \hbr$):
\begin{eqnarray}
     \hat{H}_{exact} &=& \frac{1}{2M}(\hat{\bm{K}}+e\bB\times \hbr)^2 + \frac{1}{2\mu}(\hat{\bm{p}}_r+\frac{e}{2}\bB\times \hbR + \frac{\mu}{\mu'}\frac{e}{2}\bB\times \hbr)^2+\hat{V}(\hbr)
\end{eqnarray}
Finally, to fully separate the center of mass coordinates, we further apply a unitary transform  $\hat{U} = \exp{[-\frac{ie}{2\hbar}(\bB\times \hbR)\cdot \hbr]}$ to the Hamiltonian:
\begin{eqnarray}
    \hat{H}_{exact}' &=& \hat{U}^{-1}\hat{H}\hat{U}\\
     &=& \frac{1}{2M}(\hbP_R+e\bB\times \hbr)^2 + \frac{1}{2\mu}(\hbp_r+ \frac{\mu}{\mu'}\frac{e}{2}\bB\times \hbr)^2+\hat{V}(\hbr) \label{eq:exactH}
\end{eqnarray}
In the end, $\hbP_R$ is a constant of motion for this Hamiltonian; notice that  $\hbP_R$ is the unitary transform of the pseudomomentum (i.e. the dressed pseudomomentum):
\begin{eqnarray}
    \hat{\bm{K}}' = \hat{U}^{-1}\hat{\bm{K}}\hat{U} = \hat{U}^{-1}(\hbP_R -\frac{e}{2}\bB\times \hbr)\hat{U} = \hbP_R 
\end{eqnarray}
The other quantity conserved by this Hamiltonian is the canonical angular momentum:
\begin{eqnarray}
    \hL^z = \hat{U}^{-1}(\hL_{R}^z + \hL_r^z)\hat{U} = \hL_{R}^z + \hL_r^z
\end{eqnarray}
\subsection{Phase-Space Hamiltonian}
Next, let us compare  the result above with the phase-space Hamiltonian. For the hydrogen atom, the phase-space Hamiltonian  takes the form (setting $q^{\it{eff}} = 0$): 
\begin{eqnarray}
   \hat{H}_{PS} &=& \frac{\bP_n^2}{2M_n}+\frac{\hat{\bm{\Gamma}}_e^2}{2M_n}-\frac{\bP_n\cdot\hat{\bm{\Gamma}}_e}{M_n} + \hat{H}_{el} 
\end{eqnarray}
where $\hat{\bm{\Gamma}}_e = \hat{\bm{k}}_e^n = \hat{\bm{\pi}}_e -e\bB\times (\hbr_e-\bR_n)$. 
After collecting all the $\hat{\bm{\pi}}_e$ terms and completing the square,  we find:
\begin{eqnarray}
    \hat{H}_{PS} &=& \frac{1}{2M}\Big[{\bP}_n +e\bB\times (\hat{\bm{r}}_e-\bR_n) \Big]^2 
    \nonumber\\&&+\frac{1}{2\mu}\Big[\hat{\bm{\pi}}_e -\frac{\mu}{M_n}e \bB\times (\hbr_e-\bR_n)-\frac{\mu}{M_n}\bP_n\Big]^2 + \hat{V}(\bm{\hat{r}}_e-\bR_n) \\
     &=& \frac{1}{2M}\Big[{\bP}_n +e\bB\times (\hbr_e-\bR_n) \Big]^2 
    \nonumber\\&&+\frac{1}{2\mu}\Big[\hbp_e +\frac{e}{2}\Big(\frac{M_n-m_e}{M_n+m_e}\Big) \bB\times (\hbr_e-\bR_n)-\frac{\mu}{M_n}\bP_n+\frac{3m_e-M_n}{M_n+m_e}e\bB\times \bR_n\Big]^2  \nonumber \\
    &&  + \hat{V}(\hbr_e-\bR_n)
\end{eqnarray}
At this point,  we change coordinates from $\hbr_e$ to $\hbr \equiv \hbr_e-\bR_n$:
\begin{eqnarray}
    \hat{H}_{PS} &=& \frac{1}{2M}\Big[\bP_n +e\bB\times \hbr \Big]^2 
    \nonumber\\&&+\frac{1}{2\mu}\Big[\hbp_e +\frac{e}{2}\Big(\frac{M_n-m_e}{M_n+m_e}\Big) \bB\times \hbr-\frac{\mu}{M_n}\bP_n+\frac{3m_e-M_n}{M_n+m_e}e\bB\times \bR_n\Big]^2 \nonumber + \hat{V}(\hbr)\nonumber\\
\end{eqnarray}
Finally, if we apply a unitary transform with $\hat{U}=\exp{\Big((\frac{\mu}{M_n}\bP_n-\frac{3m_e-M_n}{M_n+m_e}e\bB\times \bR_n)\cdot \hbr\Big)}$, we recover a form of phase-space Hamiltonian that strongly resembles  Eq. \ref{eq:exactH} above:
\begin{eqnarray}
    \hat{H}_{PS} &=& \frac{1}{2M}\Big[{\bP}_n +e\bB\times \hbr \Big]^2 
    +\frac{1}{2\mu}\Big[\hbp_e +\frac{e}{2}\Big(\frac{M_n-m_e}{M_n+m_e}\Big) \bB\times \hbr\Big]^2 + \hat{V}(\hbr)\label{eq:HatomPS}
\end{eqnarray}
In fact, by comparing Eq. \ref{eq:HatomPS} versus Eq. \ref{eq:exactH},  we find that our phase-space Hamiltonian is identical to an exact treatment, provided that the nuclear momentum $\bP_n$ in phase-space Hamiltonian corresponds to the center of mass momentum $\hbP_R$ and the electronic momentum $\hbp_e$ corresponds to the internal coordinate momentum $\hbp_r$.   

As a side note, if we  diagonalize the Hamiltonian in Eq. \ref{eq:HatomPS} and  calculate an eigenstate $\ket{\Phi}$ and evaluate  $E_{PS}(\bR_n,\bP_n) = \left< \Phi \middle| \hat{H}_{PS} \middle| \Phi \right>$, dynamics along an eigenstate will indeed conserve 
$P_{nX},P_{nY},P_{nZ},$ and $XP_{nY}-YP_{nX}$ (as well as the energy).

\section{Discussion and Conclusions}\label{conclude}

In Eqs. \ref{eq:PSH},  \ref{eq:gammatotal}, \ref{eq:etf} and \ref{eq:final_erf} above, we have presented above a phase-space electronic Hamiltonian that offers us the opportunity to go beyond standard BO theory and perform electronic structure calculations in a moving nuclear frame. While there is no unique phase-space electronic Hamiltonian, the proposed Hamiltonian satisfies four key target properties. First, dynamics along the resulting eigenstates conserves the total pseudomomentum and canonical angular momentum (Eqs. \ref{eq:pseudomom2} and \ref{eq:Ltot}). Second, the phase-space space energy $E_{PS}$ does not depend on $\bG$ (Eq. \ref{eq:eps_g_grad}), and therefore all of the dynamics will not depend on $\bG$. For a longer discussion of this gauge degree of freedom, see Paper II\cite{bhati2024paper2}.
Third, our proposed approach reduces to the exact answer for the hydrogen atom by setting $q^{\textit{eff}} = 0$ (Sec. \ref{results}). Fourth, in the limit of a  system with no electrons, the dynamics are exact nuclear dynamics simply by setting $q^{\textit{eff}} = q_I$. All of these attributes give us hope that we have found here a meaningful way to treat electronic structure in a magnetic field that goes beyond BO theory.

That being said, the list above is
clearly incomplete and there are several more sanity checks that must follow (before we compare against experiment). First, one would like to investigate and benchmark the vibrational energies predicted by BO (in a similar fashion to the calculation performed in Ref. \citenum{bian2024phase}). Focusing on the ground state, we know that because $E_{PS}(\bX,\bP)$ is a function of both nuclear position and momentum, generating vibrational energies will require a Wigner transformation (as in Ref. \citenum{bian2024phase}).
Second, beyond the conservation of the total pseudomomentum and angular momentum, one would like to check whether the current algorithm can produce a meaningful electronic pseudomomentum and angular momentum (in a similar fashion to Ref. \citenum{coraline_pssh}). Note that, in order to perform the two benchmarks above we will need to run dynamics on larger systems, which will necessitate the introduction of  Gauge Including Atomic Orbitals (GIAOs) for practical calculations\cite{pulay2007shielding,helgaker1991giaome,andrewtealegiao}. This subject will be taken up in Paper II of this series.
Third, it remains to check how sensitive our results will be to the choice of $q^{\textit{eff}}$. In Ref. \citenum{tao2024basis}, found that our results were not very sensitive to the choice of $\hat{\Theta}_I$, provided we had reasonably sharp barriers based on reasonable estimates of atomic radii (e.g. van der waal radii).  More generally, one could imagine using Hirshfeld charge densities\cite{hirshfeld1977bonded} to generate $\hat{\Theta}_I$ and $q^{\textit{eff}}$.  More generally, one must also wonder how accurate is the approximation in Eq. \ref{eq:approx_joe} and whether one can do better than our definition of $q^{\textit{eff}}$.
Indeed, there are many basic questions that should  be checked -- both analytically and numerically -- in order to find the optimal phase space electronic Hamiltonian approach.

All of these caveats aside, if we look forward, generating a meaningful phase-space electronic Hamiltonian in the presence of a magnetic field offers a large and  exciting list of future research projects for exploration, especially if we can incorporate spin in the future.  
At the top of this list, we would like to understand how a magnetic field affects coupled nonadiabatic nuclear-electronic dynamics  at a curve crossing. As Steiner pointed out many years ago\cite{steiner1989}, there is a long list of magnetic field effects that cannot be modeled numerically or often explained at all, but given the weakness of a typical magnetic field, the usual ansatz is that these field effects have dynamical origins. Our ansatz is therefore that many such effects can be explained by properly conserving pseudomomentum and angular momentum, and the effect of small changes is often amplified at a curve crossing. More generally, one must also wonder whether the present approach will offer us new insight into electron transfer and magnetoresistance\cite{magnetoresistance_review}, especially in organic materials, where the role of phonons in modulating charge transfer is clearly important\cite{OMAR}. There may well also be implications for the chiral induced spin selectivity effect.\cite{doi:10.1021/acs.accounts.0c00485,doi:10.1021/jz300793y} Finally, there is also strong possibility that we can use the approach here to simulate the Einstein-de Haas effect\cite{edhass1,edhass2,ganzhorn2016quantum,lemesheko:2015:prl} through the direct coupling of electronic spin angular momentum to nuclear angular momentum in Eq. \ref{eq:final_erf_s}. Overall, if the present approach can be successfully verified, one can imagine an enormous number of future applications with large impact in chemical physics.

\section{Acknowledgments}
This work has been supported by the U.S. Department of Energy, Office of Science, Office of Basic Energy Sciences, under Award no. DE-SC0025393.

\appendix
\section{Appendix: The Five Conserved Quantities in a Uniform Magnetic Field}\label{conserve_proof}

For this Appendix only, we will consider a set of charges and not distinguish between electrons and nuclei. The Hamiltonian for a collection of particles with charges $Q_I$ in a magnetic field can be written as the sum of the kinetic energy and the interaction energy between them:
\begin{eqnarray}
    \hat{H} &=& \sum_I\frac{(\hbP_I-Q_IA(\hbX_I))^2}{2M_I} + \sum_{IJ}\frac{1}{4\pi\epsilon_0}\frac{Q_IQ_J}{|\hat{\bX}_I-\hat{\bX}_J|} 
\end{eqnarray}
Here, for a hydrogen atom, $Q = +e$.  For an electron, $Q=-e$.  To find the constants of motion for this Hamiltonian, the following commutators will be helpful:
\begin{align}
    [\hat{\Pi}_{I\alpha}, \hat{\Pi}_{J\beta}] = i\hbar \sum_{I\gamma}\epsilon_{\alpha \beta \gamma} Q_I B_\gamma \delta_{IJ}\;\;\;\;\;\;\text{and}\;\;\;\;\;\;
    [\hat{X}_{I\alpha},\hat{\Pi}_{J\beta}] = i\hbar\delta_{\alpha \beta}\delta_{IJ}
\end{align}
 The equations of motion for this Hamiltonian can be obtained by taking the time derivative of the operator:
\begin{eqnarray}
    \dot{X}_{I\alpha} = \frac{i}{\hbar}[\hat{H},\hat{X}_{I\alpha}]
   = \frac{i}{\hbar}\frac{2\hat{\Pi}_{I\alpha}}{2M_I}(-i\hbar) =  \frac{\hat{\Pi}_{I\alpha}}{M_I}
\end{eqnarray}
Therefore, to find the conserved quantities we can start by taking time derivative of the kinetic momentum in the $x-$direction:  
\begin{eqnarray}
    \dot{\Pi}_{IX} &=& \frac{i}{\hbar} [\hat{H},\hat{\Pi}_{IX}] \\
    &=& \sum_J\frac{i}{\hbar} \Big[\frac{\hat{\Pi}_{JY}^2}{2M_J},\hat{\Pi}_{IX}\Big] + \frac{i}{\hbar}\Big[\hat{V}, \hat{\Pi}_{IX} \Big]\\
    &=& \sum_J \frac{i}{\hbar} \frac{1}{2M_J}(-2i\hbar QB_z\hat{\Pi}_{JY}\delta_{IJ}) + \frac{i}{\hbar}[\hat{V}, \hat{P}_{IX} ]\\
    &=& QB_z\frac{\hat{\Pi}_{IY}}{M_I}\\
    &=& QB_z\dot{Y}_I
\end{eqnarray}
Rearranging the equation above shows that the pseudomomentum in the $x-$direction is conserved.
\begin{eqnarray}
    \sum_I\Big(\dot{\Pi}_{Ix}-Q_IB_z\dot{Y}_I\Big) &=& 0\\
    \frac{d}{dt}\sum_I\Big(\hat{\Pi}_{Ix}-Q_IB_z\hat{Y}_I\Big) &=& 0
\end{eqnarray}
Similarly, the pseudomomentum in the $y$ and $z$ directions can also be shown to be conserved:
\begin{eqnarray}    \sum_I\Big(\hat{\Pi}_{Iy}+eQ_IB_z\hat{X}_I\Big) &=& \text{const.}\\
\sum_I \hat{\Pi}_{Iz} &=& \text{const.} 
\end{eqnarray}
We can similarly take the gradient of the  angular momentum,
\begin{eqnarray}
    \dot{L}_{Iz}  &=& \frac{i}{\hbar}[\hat{H}, \hat{L}_{Iz}] \\
    &=& \frac{i}{\hbar}[\hat{T}_n, \hat{X}_I \hat{\Pi}_{IY}-\hat{Y}_I\hat{\Pi}_{IX}] + \frac{i}{\hbar}[\hat{V},\hat{L}_{Iz}]\\
    &=& \frac{i}{\hbar} \Big([\hat{T}_n, \hat{X}_I] \hat{\Pi}_{IY}+ [\hat{T}_n, \hat{\Pi}_{IY}]\hat{X}_I -[\hat{T}_n, \hat{Y}_I]\hat{\Pi}_{IX}-[\hat{T}_n, \hat{\Pi}_{IX}]\hat{Y}_I\Big)\\
    &=& \frac{i}{\hbar}\sum_J\frac{1}{2M_J} \Big((-i\hbar\hat{\Pi}_{JX}\delta_{IJ})\hat{\Pi}_{IY}+ (2i\hbar QB_z\hat{\Pi}_{JX}\delta_{IJ})\hat{X}_I \\&&-(-i\hbar\hat{\Pi}_{JY}\delta_{IJ})\hat{\Pi}_{IX}-(-2i\hbar QB_z\hat{\Pi}_{JY}\delta_{IJ})\hat{Y}_I \Big) \\
    &=& -QB_z(\dot{X}_I\hat{X}_I+\dot{Y}_I\hat{Y}_I)
\end{eqnarray}
Rearranging the above terms shows that  the total canonical angular momentum is conserved.
\begin{eqnarray}
\frac{d}{dt}\sum_I\Big( \hat{L}_{Iz} + \frac{QB_z}{2}(\hat{X}_I^2+\hat{Y}_I^2)\Big) &=& 0 
\end{eqnarray}
Lastly, since the magnetic fields do not perform work on the system, the energy of the system is also conserved (of course).

\bibliography{Ref}

\providecommand{\latin}[1]{#1}
\makeatletter
\providecommand{\doi}
  {\begingroup\let\do\@makeother\dospecials
  \catcode`\{=1 \catcode`\}=2 \doi@aux}
\providecommand{\doi@aux}[1]{\endgroup\texttt{#1}}
\makeatother
\providecommand*\mcitethebibliography{\thebibliography}
\csname @ifundefined\endcsname{endmcitethebibliography}  {\let\endmcitethebibliography\endthebibliography}{}
\begin{mcitethebibliography}{59}
\providecommand*\natexlab[1]{#1}
\providecommand*\mciteSetBstSublistMode[1]{}
\providecommand*\mciteSetBstMaxWidthForm[2]{}
\providecommand*\mciteBstWouldAddEndPuncttrue
  {\def\EndOfBibitem{\unskip.}}
\providecommand*\mciteBstWouldAddEndPunctfalse
  {\let\EndOfBibitem\relax}
\providecommand*\mciteSetBstMidEndSepPunct[3]{}
\providecommand*\mciteSetBstSublistLabelBeginEnd[3]{}
\providecommand*\EndOfBibitem{}
\mciteSetBstSublistMode{f}
\mciteSetBstMaxWidthForm{subitem}{(\alph{mcitesubitemcount})}
\mciteSetBstSublistLabelBeginEnd
  {\mcitemaxwidthsubitemform\space}
  {\relax}
  {\relax}

\bibitem[Slichter(2013)]{nmrbook}
Slichter,~C.~P. \emph{Principles of magnetic resonance}; Springer Science \& Business Media, 2013; Vol.~1\relax
\mciteBstWouldAddEndPuncttrue
\mciteSetBstMidEndSepPunct{\mcitedefaultmidpunct}
{\mcitedefaultendpunct}{\mcitedefaultseppunct}\relax
\EndOfBibitem
\bibitem[Steiner and Ulrich(1989)Steiner, and Ulrich]{steiner1989}
Steiner,~U.~E.; Ulrich,~T. Magnetic field effects in chemical kinetics and related phenomena. \emph{Chemical Reviews} \textbf{1989}, \emph{89}, 51--147\relax
\mciteBstWouldAddEndPuncttrue
\mciteSetBstMidEndSepPunct{\mcitedefaultmidpunct}
{\mcitedefaultendpunct}{\mcitedefaultseppunct}\relax
\EndOfBibitem
\bibitem[Rodgers(2009)]{rodgers2009magnetic}
Rodgers,~C.~T. Magnetic field effects in chemical systems. \emph{Pure and Applied Chemistry} \textbf{2009}, \emph{81}, 19--43\relax
\mciteBstWouldAddEndPuncttrue
\mciteSetBstMidEndSepPunct{\mcitedefaultmidpunct}
{\mcitedefaultendpunct}{\mcitedefaultseppunct}\relax
\EndOfBibitem
\bibitem[Rodgers and Hore(2009)Rodgers, and Hore]{pnas_magnetoreception}
Rodgers,~C.~T.; Hore,~P.~J. Chemical magnetoreception in birds: The radical pair mechanism. \emph{Proceedings of the National Academy of Sciences} \textbf{2009}, \emph{106}, 353--360\relax
\mciteBstWouldAddEndPuncttrue
\mciteSetBstMidEndSepPunct{\mcitedefaultmidpunct}
{\mcitedefaultendpunct}{\mcitedefaultseppunct}\relax
\EndOfBibitem
\bibitem[Luo \latin{et~al.}(2024)Luo, Benjamin, Gerhards, Hogben, and Hore]{radicalpair}
Luo,~J.; Benjamin,~P.; Gerhards,~L.; Hogben,~H.~J.; Hore,~P.~J. Orientation of birds in radiofrequency fields in the absence of the Earth’s magnetic field: a possible test for the radical pair mechanism of magnetoreception. \emph{Journal of The Royal Society Interface} \textbf{2024}, \emph{21}, 20240133\relax
\mciteBstWouldAddEndPuncttrue
\mciteSetBstMidEndSepPunct{\mcitedefaultmidpunct}
{\mcitedefaultendpunct}{\mcitedefaultseppunct}\relax
\EndOfBibitem
\bibitem[Thompson \latin{et~al.}(2004)Thompson, Timmel, and Henbest]{spinweakmagneticfield}
Thompson,~J. M.~T.; Timmel,~C.~R.; Henbest,~K.~B. A study of spin chemistry in weak magnetic fields. \emph{Philosophical Transactions of the Royal Society of London. Series A: Mathematical, Physical and Engineering Sciences} \textbf{2004}, \emph{362}, 2573--2589\relax
\mciteBstWouldAddEndPuncttrue
\mciteSetBstMidEndSepPunct{\mcitedefaultmidpunct}
{\mcitedefaultendpunct}{\mcitedefaultseppunct}\relax
\EndOfBibitem
\bibitem[Stopkowicz \latin{et~al.}(2015)Stopkowicz, Gauss, Lange, Tellgren, and Helgaker]{stopkowicz2015coupled}
Stopkowicz,~S.; Gauss,~J.; Lange,~K.~K.; Tellgren,~E.~I.; Helgaker,~T. Coupled-cluster theory for atoms and molecules in strong magnetic fields. \emph{The Journal of Chemical Physics} \textbf{2015}, \emph{143}\relax
\mciteBstWouldAddEndPuncttrue
\mciteSetBstMidEndSepPunct{\mcitedefaultmidpunct}
{\mcitedefaultendpunct}{\mcitedefaultseppunct}\relax
\EndOfBibitem
\bibitem[Schmelcher and Cederbaum(1990)Schmelcher, and Cederbaum]{cederbaum1990}
Schmelcher,~P.; Cederbaum,~L.~S. Crossings of potential-energy surfaces in a magnetic field. \emph{Phys. Rev. A} \textbf{1990}, \emph{41}, 4936--4943\relax
\mciteBstWouldAddEndPuncttrue
\mciteSetBstMidEndSepPunct{\mcitedefaultmidpunct}
{\mcitedefaultendpunct}{\mcitedefaultseppunct}\relax
\EndOfBibitem
\bibitem[Brigham and Wadehra(1987)Brigham, and Wadehra]{brigham1987}
Brigham,~D.~R.; Wadehra,~J. The hydrogen molecular ion in an arbitrary homogeneous magnetic field. \emph{Astrophysical Journal, Part 1 (ISSN 0004-637X), vol. 317, June 15, 1987, p. 865-876.} \textbf{1987}, \emph{317}, 865--876\relax
\mciteBstWouldAddEndPuncttrue
\mciteSetBstMidEndSepPunct{\mcitedefaultmidpunct}
{\mcitedefaultendpunct}{\mcitedefaultseppunct}\relax
\EndOfBibitem
\bibitem[Schmelcher and Cederbaum(1989)Schmelcher, and Cederbaum]{cederbaum1989}
Schmelcher,~P.; Cederbaum,~L.~S. Approximate constant of motion for molecular ions in a magnetic field. \emph{Phys. Rev. A} \textbf{1989}, \emph{40}, 3515--3523\relax
\mciteBstWouldAddEndPuncttrue
\mciteSetBstMidEndSepPunct{\mcitedefaultmidpunct}
{\mcitedefaultendpunct}{\mcitedefaultseppunct}\relax
\EndOfBibitem
\bibitem[Lai(2001)]{lai2001matter}
Lai,~D. Matter in strong magnetic fields. \emph{Reviews of Modern Physics} \textbf{2001}, \emph{73}, 629\relax
\mciteBstWouldAddEndPuncttrue
\mciteSetBstMidEndSepPunct{\mcitedefaultmidpunct}
{\mcitedefaultendpunct}{\mcitedefaultseppunct}\relax
\EndOfBibitem
\bibitem[Schmelcher \latin{et~al.}(1988)Schmelcher, Cederbaum, and Meyer]{cederbaum1988}
Schmelcher,~P.; Cederbaum,~L.~S.; Meyer,~H.-D. Electronic and nuclear motion and their couplings in the presence of a magnetic field. \emph{Phys. Rev. A} \textbf{1988}, \emph{38}, 6066--6079\relax
\mciteBstWouldAddEndPuncttrue
\mciteSetBstMidEndSepPunct{\mcitedefaultmidpunct}
{\mcitedefaultendpunct}{\mcitedefaultseppunct}\relax
\EndOfBibitem
\bibitem[Yin and Alden~Mead(1994)Yin, and Alden~Mead]{yin1994magnetic}
Yin,~L.; Alden~Mead,~C. Magnetic screening of nuclei by electrons as an effect of geometric vector potential. \emph{The Journal of chemical physics} \textbf{1994}, \emph{100}, 8125--8131\relax
\mciteBstWouldAddEndPuncttrue
\mciteSetBstMidEndSepPunct{\mcitedefaultmidpunct}
{\mcitedefaultendpunct}{\mcitedefaultseppunct}\relax
\EndOfBibitem
\bibitem[Schmelcher \latin{et~al.}(1988)Schmelcher, Cederbaum, and Meyer]{Schmelcher_1988}
Schmelcher,~P.; Cederbaum,~L.~S.; Meyer,~H.~D. On the validity of the Born-Oppenheimer approximation in magnetic fields. \emph{Journal of Physics B: Atomic, Molecular and Optical Physics} \textbf{1988}, \emph{21}, L445\relax
\mciteBstWouldAddEndPuncttrue
\mciteSetBstMidEndSepPunct{\mcitedefaultmidpunct}
{\mcitedefaultendpunct}{\mcitedefaultseppunct}\relax
\EndOfBibitem
\bibitem[Malhado \latin{et~al.}(2014)Malhado, Bearpark, and Hynes]{hynes:ci_fssh:review}
Malhado,~J.~P.; Bearpark,~M.~J.; Hynes,~J.~T. Non-adiabatic dynamics close to conical intersections and the surface hopping perspective. \emph{Frontiers in Chemistry} \textbf{2014}, \emph{2}\relax
\mciteBstWouldAddEndPuncttrue
\mciteSetBstMidEndSepPunct{\mcitedefaultmidpunct}
{\mcitedefaultendpunct}{\mcitedefaultseppunct}\relax
\EndOfBibitem
\bibitem[Tully(1990)]{tullysurfacehopping}
Tully,~J.~C. {Molecular dynamics with electronic transitions}. \emph{The Journal of Chemical Physics} \textbf{1990}, \emph{93}, 1061--1071\relax
\mciteBstWouldAddEndPuncttrue
\mciteSetBstMidEndSepPunct{\mcitedefaultmidpunct}
{\mcitedefaultendpunct}{\mcitedefaultseppunct}\relax
\EndOfBibitem
\bibitem[Tao \latin{et~al.}(2024)Tao, Qiu, Bhati, Bian, Duston, Rawlinson, Littlejohn, and Subotnik]{coraline_pssh}
Tao,~Z.; Qiu,~T.; Bhati,~M.; Bian,~X.; Duston,~T.; Rawlinson,~J.; Littlejohn,~R.~G.; Subotnik,~J.~E. {Practical phase-space electronic Hamiltonians for ab-initio dynamics}. \emph{The Journal of Chemical Physics} \textbf{2024}, \emph{160}, 124101\relax
\mciteBstWouldAddEndPuncttrue
\mciteSetBstMidEndSepPunct{\mcitedefaultmidpunct}
{\mcitedefaultendpunct}{\mcitedefaultseppunct}\relax
\EndOfBibitem
\bibitem[Littlejohn \latin{et~al.}(2023)Littlejohn, Rawlinson, and Subotnik]{Littlejohn2023}
Littlejohn,~R.; Rawlinson,~J.; Subotnik,~J. Representation and conservation of angular momentum in the Born–Oppenheimer theory of polyatomic molecules. \emph{The Journal of Chemical Physics} \textbf{2023}, \emph{158}, 104302\relax
\mciteBstWouldAddEndPuncttrue
\mciteSetBstMidEndSepPunct{\mcitedefaultmidpunct}
{\mcitedefaultendpunct}{\mcitedefaultseppunct}\relax
\EndOfBibitem
\bibitem[Bian \latin{et~al.}(2023)Bian, Tao, Wu, Rawlinson, Littlejohn, and Subotnik]{Bian2023}
Bian,~X.; Tao,~Z.; Wu,~Y.; Rawlinson,~J.; Littlejohn,~R.~G.; Subotnik,~J.~E. Total angular momentum conservation in ab initio Born-Oppenheimer molecular dynamics. \emph{Phys. Rev. B} \textbf{2023}, \emph{108}, L220304\relax
\mciteBstWouldAddEndPuncttrue
\mciteSetBstMidEndSepPunct{\mcitedefaultmidpunct}
{\mcitedefaultendpunct}{\mcitedefaultseppunct}\relax
\EndOfBibitem
\bibitem[Mead and Truhlar(1979)Mead, and Truhlar]{mead1979determination}
Mead,~C.~A.; Truhlar,~D.~G. On the determination of Born--Oppenheimer nuclear motion wave functions including complications due to conical intersections and identical nuclei. \emph{The Journal of Chemical Physics} \textbf{1979}, \emph{70}, 2284--2296\relax
\mciteBstWouldAddEndPuncttrue
\mciteSetBstMidEndSepPunct{\mcitedefaultmidpunct}
{\mcitedefaultendpunct}{\mcitedefaultseppunct}\relax
\EndOfBibitem
\bibitem[Berry(1984)]{berry1984quantal}
Berry,~M.~V. Quantal phase factors accompanying adiabatic changes. \emph{Proceedings of the Royal Society of London. A. Mathematical and Physical Sciences} \textbf{1984}, \emph{392}, 45--57\relax
\mciteBstWouldAddEndPuncttrue
\mciteSetBstMidEndSepPunct{\mcitedefaultmidpunct}
{\mcitedefaultendpunct}{\mcitedefaultseppunct}\relax
\EndOfBibitem
\bibitem[Peters \latin{et~al.}(2022)Peters, Culpitt, Tellgren, and Helgaker]{magnetictrans}
Peters,~L. D.~M.; Culpitt,~T.; Tellgren,~E.~I.; Helgaker,~T. {Magnetic-translational sum rule and approximate models of the molecular Berry curvature}. \emph{The Journal of Chemical Physics} \textbf{2022}, \emph{157}, 134108\relax
\mciteBstWouldAddEndPuncttrue
\mciteSetBstMidEndSepPunct{\mcitedefaultmidpunct}
{\mcitedefaultendpunct}{\mcitedefaultseppunct}\relax
\EndOfBibitem
\bibitem[Culpitt \latin{et~al.}(2022)Culpitt, Peters, Tellgren, and Helgaker]{Culpitt2022}
Culpitt,~T.; Peters,~L. D.~M.; Tellgren,~E.~I.; Helgaker,~T. Analytic calculation of the Berry curvature and diagonal Born–Oppenheimer correction for molecular systems in uniform magnetic fields. \emph{J. Chem. Phys.} \textbf{2022}, \emph{156}, 044121\relax
\mciteBstWouldAddEndPuncttrue
\mciteSetBstMidEndSepPunct{\mcitedefaultmidpunct}
{\mcitedefaultendpunct}{\mcitedefaultseppunct}\relax
\EndOfBibitem
\bibitem[Bian \latin{et~al.}(2021)Bian, Wu, Teh, Zhou, Chen, and Subotnik]{Bian2021}
Bian,~X.; Wu,~Y.; Teh,~H.-H.; Zhou,~Z.; Chen,~H.-T.; Subotnik,~J.~E. {Modeling nonadiabatic dynamics with degenerate electronic states, intersystem crossing, and spin separation: A key goal for chemical physics}. \emph{The Journal of Chemical Physics} \textbf{2021}, \emph{154}, 110901\relax
\mciteBstWouldAddEndPuncttrue
\mciteSetBstMidEndSepPunct{\mcitedefaultmidpunct}
{\mcitedefaultendpunct}{\mcitedefaultseppunct}\relax
\EndOfBibitem
\bibitem[cit()]{citation-key}
As a side note, FSSH dynamics using BO states in the presence of a magnetic field is complicated by the existence of complex derivative coupling values, which makes the direction of momentum scaling ambiguous; this problem can be partially ameliorated using a phase space approach.\relax
\mciteBstWouldAddEndPunctfalse
\mciteSetBstMidEndSepPunct{\mcitedefaultmidpunct}
{}{\mcitedefaultseppunct}\relax
\EndOfBibitem
\bibitem[Qiu \latin{et~al.}(2024)Qiu, Bhati, Tao, Bian, Rawlinson, Littlejohn, and Subotnik]{tian_erf}
Qiu,~T.; Bhati,~M.; Tao,~Z.; Bian,~X.; Rawlinson,~J.; Littlejohn,~R.~G.; Subotnik,~J.~E. {A simple one-electron expression for electron rotational factors}. \emph{The Journal of Chemical Physics} \textbf{2024}, \emph{160}, 124102\relax
\mciteBstWouldAddEndPuncttrue
\mciteSetBstMidEndSepPunct{\mcitedefaultmidpunct}
{\mcitedefaultendpunct}{\mcitedefaultseppunct}\relax
\EndOfBibitem
\bibitem[Tao \latin{et~al.}(2024)Tao, Qiu, Bian, and Subotnik]{tao2024basis}
Tao,~Z.; Qiu,~T.; Bian,~X.; Subotnik,~J.~E. A Basis-Free Phase Space Electronic Hamiltonian That Recovers Beyond Born-Oppenheimer Electronic Momentum and Current Density. \emph{arXiv preprint arXiv:2407.16918} \textbf{2024}, \relax
\mciteBstWouldAddEndPunctfalse
\mciteSetBstMidEndSepPunct{\mcitedefaultmidpunct}
{}{\mcitedefaultseppunct}\relax
\EndOfBibitem
\bibitem[Duston \latin{et~al.}(2024)Duston, Tao, Bian, Bhati, Rawlinson, Littlejohn, Pei, Shao, and Subotnik]{duston2024phase}
Duston,~T.; Tao,~Z.; Bian,~X.; Bhati,~M.; Rawlinson,~J.; Littlejohn,~R.~G.; Pei,~Z.; Shao,~Y.; Subotnik,~J.~E. A Phase-Space Electronic Hamiltonian For Vibrational Circular Dichroism. \emph{Journal of Chemical Theory and Computation} \textbf{2024}, \emph{20}, 7904--7921, PMID: 39226223\relax
\mciteBstWouldAddEndPuncttrue
\mciteSetBstMidEndSepPunct{\mcitedefaultmidpunct}
{\mcitedefaultendpunct}{\mcitedefaultseppunct}\relax
\EndOfBibitem
\bibitem[Wu \latin{et~al.}(2024)Wu, Rawlinson, Littlejohn, and Subotnik]{wu2024linear}
Wu,~Y.; Rawlinson,~J.; Littlejohn,~R.~G.; Subotnik,~J.~E. Linear and angular momentum conservation in surface hopping methods. \emph{The Journal of Chemical Physics} \textbf{2024}, \emph{160}\relax
\mciteBstWouldAddEndPuncttrue
\mciteSetBstMidEndSepPunct{\mcitedefaultmidpunct}
{\mcitedefaultendpunct}{\mcitedefaultseppunct}\relax
\EndOfBibitem
\bibitem[Shenvi(2009)]{shenvi:2009:jcp_pssh}
Shenvi,~N. Phase-space surface hopping: Nonadiabatic dynamics in a superadiabatic basis. \emph{Journal of Chemical Physics} \textbf{2009}, \emph{130}, 124117\relax
\mciteBstWouldAddEndPuncttrue
\mciteSetBstMidEndSepPunct{\mcitedefaultmidpunct}
{\mcitedefaultendpunct}{\mcitedefaultseppunct}\relax
\EndOfBibitem
\bibitem[Peters \latin{et~al.}(2021)Peters, Culpitt, Monzel, Tellgren, and Helgaker]{peters2021ab}
Peters,~L.~D.; Culpitt,~T.; Monzel,~L.; Tellgren,~E.~I.; Helgaker,~T. Ab Initio molecular dynamics with screened Lorentz forces. II. Efficient propagators and rovibrational spectra in strong magnetic fields. \emph{The Journal of Chemical Physics} \textbf{2021}, \emph{155}\relax
\mciteBstWouldAddEndPuncttrue
\mciteSetBstMidEndSepPunct{\mcitedefaultmidpunct}
{\mcitedefaultendpunct}{\mcitedefaultseppunct}\relax
\EndOfBibitem
\bibitem[Abedi \latin{et~al.}(2010)Abedi, Maitra, and Gross]{PhysRevLett.105.123002}
Abedi,~A.; Maitra,~N.~T.; Gross,~E. K.~U. Exact Factorization of the Time-Dependent Electron-Nuclear Wave Function. \emph{Phys. Rev. Lett.} \textbf{2010}, \emph{105}, 123002\relax
\mciteBstWouldAddEndPuncttrue
\mciteSetBstMidEndSepPunct{\mcitedefaultmidpunct}
{\mcitedefaultendpunct}{\mcitedefaultseppunct}\relax
\EndOfBibitem
\bibitem[Scherrer \latin{et~al.}(2015)Scherrer, Agostini, Sebastiani, Gross, and Vuilleumier]{scherrer2015nuclear}
Scherrer,~A.; Agostini,~F.; Sebastiani,~D.; Gross,~E.; Vuilleumier,~R. Nuclear velocity perturbation theory for vibrational circular dichroism: An approach based on the exact factorization of the electron-nuclear wave function. \emph{The Journal of chemical physics} \textbf{2015}, \emph{143}\relax
\mciteBstWouldAddEndPuncttrue
\mciteSetBstMidEndSepPunct{\mcitedefaultmidpunct}
{\mcitedefaultendpunct}{\mcitedefaultseppunct}\relax
\EndOfBibitem
\bibitem[Li \latin{et~al.}(2022)Li, Requist, and Gross]{PhysRevLett.128.113001}
Li,~C.; Requist,~R.; Gross,~E. K.~U. Energy, Momentum, and Angular Momentum Transfer between Electrons and Nuclei. \emph{Phys. Rev. Lett.} \textbf{2022}, \emph{128}, 113001\relax
\mciteBstWouldAddEndPuncttrue
\mciteSetBstMidEndSepPunct{\mcitedefaultmidpunct}
{\mcitedefaultendpunct}{\mcitedefaultseppunct}\relax
\EndOfBibitem
\bibitem[Bhati \latin{et~al.}(2024)Bhati, Tao, Bian, Rawlinson, Littlejohn, and Subotnik]{bhati2024paper2}
Bhati,~M.; Tao,~Z.; Bian,~X.; Rawlinson,~J.; Littlejohn,~R.; Subotnik,~J.~E. A Phase-Space Electronic Hamiltonian for Molecules in a Static Magnetic Field II: Quantum Chemistry Calculations with Gauge Invariant Atomic Orbitals. \emph{arXiv preprint arXiv:2411.13879} \textbf{2024}, \relax
\mciteBstWouldAddEndPunctfalse
\mciteSetBstMidEndSepPunct{\mcitedefaultmidpunct}
{}{\mcitedefaultseppunct}\relax
\EndOfBibitem
\bibitem[Fatehi \latin{et~al.}(2011)Fatehi, Alguire, Shao, and Subotnik]{fatehi_etf}
Fatehi,~S.; Alguire,~E.; Shao,~Y.; Subotnik,~J.~E. {Analytic derivative couplings between configuration-interaction-singles states with built-in electron-translation factors for translational invariance}. \emph{The Journal of Chemical Physics} \textbf{2011}, \emph{135}, 234105\relax
\mciteBstWouldAddEndPuncttrue
\mciteSetBstMidEndSepPunct{\mcitedefaultmidpunct}
{\mcitedefaultendpunct}{\mcitedefaultseppunct}\relax
\EndOfBibitem
\bibitem[Schneiderman and Russek(1969)Schneiderman, and Russek]{schneiderman:1969:pr:etf}
Schneiderman,~S.~B.; Russek,~A. Velocity-Dependent Orbitals in Proton-On-Hydrogen-Atom Collisions. \emph{Physical Review} \textbf{1969}, \emph{181}, 311--321\relax
\mciteBstWouldAddEndPuncttrue
\mciteSetBstMidEndSepPunct{\mcitedefaultmidpunct}
{\mcitedefaultendpunct}{\mcitedefaultseppunct}\relax
\EndOfBibitem
\bibitem[Athavale \latin{et~al.}(2023)Athavale, Bian, Tao, Wu, Qiu, Rawlinson, Littlejohn, and Subotnik]{athavale:2023:erf}
Athavale,~V.; Bian,~X.; Tao,~Z.; Wu,~Y.; Qiu,~T.; Rawlinson,~J.; Littlejohn,~R.~G.; Subotnik,~J.~E. Surface Hopping, Electron Translation Factors, Electron Rotation Factors, Momentum Conservation, and Size Consistency. \emph{Journal of Chemical Physics} \textbf{2023}, \emph{159}, 114120, https://dx.doi.org/10.1063/5.0160965\relax
\mciteBstWouldAddEndPuncttrue
\mciteSetBstMidEndSepPunct{\mcitedefaultmidpunct}
{\mcitedefaultendpunct}{\mcitedefaultseppunct}\relax
\EndOfBibitem
\bibitem[Greenshields \latin{et~al.}(2014)Greenshields, Stamps, Franke-Arnold, and Barnett]{Greenshields2014}
Greenshields,~C.~R.; Stamps,~R.~L.; Franke-Arnold,~S.; Barnett,~S.~M. Is the Angular Momentum of an Electron Conserved in a Uniform Magnetic Field? \emph{Phys. Rev. Lett.} \textbf{2014}, \emph{113}, 240404\relax
\mciteBstWouldAddEndPuncttrue
\mciteSetBstMidEndSepPunct{\mcitedefaultmidpunct}
{\mcitedefaultendpunct}{\mcitedefaultseppunct}\relax
\EndOfBibitem
\bibitem[Johnson \latin{et~al.}(1983)Johnson, Hirschfelder, and Yang]{Johnson1983}
Johnson,~B.~R.; Hirschfelder,~J.~O.; Yang,~K.-H. Interaction of atoms, molecules, and ions with constant electric and magnetic fields. \emph{Rev. Mod. Phys.} \textbf{1983}, \emph{55}, 109--153\relax
\mciteBstWouldAddEndPuncttrue
\mciteSetBstMidEndSepPunct{\mcitedefaultmidpunct}
{\mcitedefaultendpunct}{\mcitedefaultseppunct}\relax
\EndOfBibitem
\bibitem[Bian \latin{et~al.}(2024)Bian, Wu, Qiu, Zhen, and Subotnik]{bian2024surfacehopping}
Bian,~X.; Wu,~Y.; Qiu,~T.; Zhen,~T.; Subotnik,~J.~E. A semiclassical non-adiabatic phase-space approach to molecular translations and rotations: A new picture of surface hopping and electronic inertial effects. \textbf{2024}, \relax
\mciteBstWouldAddEndPunctfalse
\mciteSetBstMidEndSepPunct{\mcitedefaultmidpunct}
{}{\mcitedefaultseppunct}\relax
\EndOfBibitem
\bibitem[Neese(2005)]{neese2005efficient}
Neese,~F. Efficient and accurate approximations to the molecular spin-orbit coupling operator and their use in molecular g-tensor calculations. \emph{The Journal of chemical physics} \textbf{2005}, \emph{122}\relax
\mciteBstWouldAddEndPuncttrue
\mciteSetBstMidEndSepPunct{\mcitedefaultmidpunct}
{\mcitedefaultendpunct}{\mcitedefaultseppunct}\relax
\EndOfBibitem
\bibitem[Tang \latin{et~al.}(2024)Tang, Sun, and Li]{tang2024exact}
Tang,~D.; Sun,~S.; Li,~X. Exact-Two-Component Complete Active Space Method with Variational Treatment of Magnetic Field and Spin--Orbit Coupling: Application to X-ray Magnetic Circular Dichroism Spectroscopy. \emph{Journal of Chemical Theory and Computation} \textbf{2024}, \relax
\mciteBstWouldAddEndPunctfalse
\mciteSetBstMidEndSepPunct{\mcitedefaultmidpunct}
{}{\mcitedefaultseppunct}\relax
\EndOfBibitem
\bibitem[Kravchenko \latin{et~al.}(1996)Kravchenko, Liberman, and Johansson]{Hexactsoln}
Kravchenko,~Y.~P.; Liberman,~M.~A.; Johansson,~B. Exact solution for a hydrogen atom in a magnetic field of arbitrary strength. \emph{Phys. Rev. A} \textbf{1996}, \emph{54}, 287--305\relax
\mciteBstWouldAddEndPuncttrue
\mciteSetBstMidEndSepPunct{\mcitedefaultmidpunct}
{\mcitedefaultendpunct}{\mcitedefaultseppunct}\relax
\EndOfBibitem
\bibitem[Schmelcher and Cederbaum(1992)Schmelcher, and Cederbaum]{schmelcher1992regularity}
Schmelcher,~P.; Cederbaum,~L. Regularity and chaos in the center of mass motion of the hydrogen atom in a magnetic field. \emph{Zeitschrift f{\"u}r Physik D Atoms, Molecules and Clusters} \textbf{1992}, \emph{24}, 311--323\relax
\mciteBstWouldAddEndPuncttrue
\mciteSetBstMidEndSepPunct{\mcitedefaultmidpunct}
{\mcitedefaultendpunct}{\mcitedefaultseppunct}\relax
\EndOfBibitem
\bibitem[Bian \latin{et~al.}(2024)Bian, Khan, Duston, Rawlinson, Littlejohn, and Subotnik]{bian2024phase}
Bian,~X.; Khan,~C.; Duston,~T.; Rawlinson,~J.; Littlejohn,~R.~G.; Subotnik,~J.~E. A phase-space view of vibrational energies without the Born-Oppenheimer framework. \emph{arXiv preprint arXiv:2407.19313} \textbf{2024}, \relax
\mciteBstWouldAddEndPunctfalse
\mciteSetBstMidEndSepPunct{\mcitedefaultmidpunct}
{}{\mcitedefaultseppunct}\relax
\EndOfBibitem
\bibitem[Pulay and Hinton(2007)Pulay, and Hinton]{pulay2007shielding}
Pulay,~P.; Hinton,~J.~F. Shielding theory: GIAO method. \emph{eMagRes} \textbf{2007}, \relax
\mciteBstWouldAddEndPunctfalse
\mciteSetBstMidEndSepPunct{\mcitedefaultmidpunct}
{}{\mcitedefaultseppunct}\relax
\EndOfBibitem
\bibitem[Helgaker and Jo/rgensen(1991)Helgaker, and Jo/rgensen]{helgaker1991giaome}
Helgaker,~T.; Jo/rgensen,~P. An electronic Hamiltonian for origin independent calculations of magnetic properties. \emph{The Journal of chemical physics} \textbf{1991}, \emph{95}, 2595--2601\relax
\mciteBstWouldAddEndPuncttrue
\mciteSetBstMidEndSepPunct{\mcitedefaultmidpunct}
{\mcitedefaultendpunct}{\mcitedefaultseppunct}\relax
\EndOfBibitem
\bibitem[Irons \latin{et~al.}(2017)Irons, Zemen, and Teale]{andrewtealegiao}
Irons,~T. J.~P.; Zemen,~J.; Teale,~A.~M. Efficient Calculation of Molecular Integrals over London Atomic Orbitals. \emph{Journal of Chemical Theory and Computation} \textbf{2017}, \emph{13}, 3636--3649, PMID: 28692291\relax
\mciteBstWouldAddEndPuncttrue
\mciteSetBstMidEndSepPunct{\mcitedefaultmidpunct}
{\mcitedefaultendpunct}{\mcitedefaultseppunct}\relax
\EndOfBibitem
\bibitem[Hirshfeld(1977)]{hirshfeld1977bonded}
Hirshfeld,~F.~L. Bonded-atom fragments for describing molecular charge densities. \emph{Theoretica chimica acta} \textbf{1977}, \emph{44}, 129--138\relax
\mciteBstWouldAddEndPuncttrue
\mciteSetBstMidEndSepPunct{\mcitedefaultmidpunct}
{\mcitedefaultendpunct}{\mcitedefaultseppunct}\relax
\EndOfBibitem
\bibitem[Gu \latin{et~al.}(2013)Gu, Zhang, Wei, Huang, Wei, and Guo]{magnetoresistance_review}
Gu,~H.; Zhang,~X.; Wei,~H.; Huang,~Y.; Wei,~S.; Guo,~Z. An overview of the magnetoresistance phenomenon in molecular systems. \emph{Chem. Soc. Rev.} \textbf{2013}, \emph{42}, 5907--5943\relax
\mciteBstWouldAddEndPuncttrue
\mciteSetBstMidEndSepPunct{\mcitedefaultmidpunct}
{\mcitedefaultendpunct}{\mcitedefaultseppunct}\relax
\EndOfBibitem
\bibitem[Gobbi and Orgiu(2017)Gobbi, and Orgiu]{OMAR}
Gobbi,~M.; Orgiu,~E. The rise of organic magnetoresistance: materials and challenges. \emph{J. Mater. Chem. C} \textbf{2017}, \emph{5}, 5572--5580\relax
\mciteBstWouldAddEndPuncttrue
\mciteSetBstMidEndSepPunct{\mcitedefaultmidpunct}
{\mcitedefaultendpunct}{\mcitedefaultseppunct}\relax
\EndOfBibitem
\bibitem[Naaman \latin{et~al.}(2020)Naaman, Paltiel, and Waldeck]{doi:10.1021/acs.accounts.0c00485}
Naaman,~R.; Paltiel,~Y.; Waldeck,~D.~H. Chiral Induced Spin Selectivity Gives a New Twist on Spin-Control in Chemistry. \emph{Accounts of Chemical Research} \textbf{2020}, \emph{53}, 2659--2667, PMID: 33044813\relax
\mciteBstWouldAddEndPuncttrue
\mciteSetBstMidEndSepPunct{\mcitedefaultmidpunct}
{\mcitedefaultendpunct}{\mcitedefaultseppunct}\relax
\EndOfBibitem
\bibitem[Naaman and Waldeck(2012)Naaman, and Waldeck]{doi:10.1021/jz300793y}
Naaman,~R.; Waldeck,~D.~H. Chiral-Induced Spin Selectivity Effect. \emph{The Journal of Physical Chemistry Letters} \textbf{2012}, \emph{3}, 2178--2187, PMID: 26295768\relax
\mciteBstWouldAddEndPuncttrue
\mciteSetBstMidEndSepPunct{\mcitedefaultmidpunct}
{\mcitedefaultendpunct}{\mcitedefaultseppunct}\relax
\EndOfBibitem
\bibitem[Richardson(1908)]{edhass1}
Richardson,~O.~W. A Mechanical Effect Accompanying Magnetization. \emph{Phys. Rev. (Series I)} \textbf{1908}, \emph{26}, 248--253\relax
\mciteBstWouldAddEndPuncttrue
\mciteSetBstMidEndSepPunct{\mcitedefaultmidpunct}
{\mcitedefaultendpunct}{\mcitedefaultseppunct}\relax
\EndOfBibitem
\bibitem[{Einstein} and {de Haas}(1915){Einstein}, and {de Haas}]{edhass2}
{Einstein},~A.; {de Haas},~W.~J. {Experimental proof of the existence of Amp{\`e}re's molecular currents}. \emph{Koninklijke Nederlandse Akademie van Wetenschappen Proceedings Series B Physical Sciences} \textbf{1915}, \emph{18}, 696--711\relax
\mciteBstWouldAddEndPuncttrue
\mciteSetBstMidEndSepPunct{\mcitedefaultmidpunct}
{\mcitedefaultendpunct}{\mcitedefaultseppunct}\relax
\EndOfBibitem
\bibitem[Ganzhorn \latin{et~al.}(2016)Ganzhorn, Klyatskaya, Ruben, and Wernsdorfer]{ganzhorn2016quantum}
Ganzhorn,~M.; Klyatskaya,~S.; Ruben,~M.; Wernsdorfer,~W. Quantum Einstein-de haas effect. \emph{Nature Communications} \textbf{2016}, \emph{7}, 11443\relax
\mciteBstWouldAddEndPuncttrue
\mciteSetBstMidEndSepPunct{\mcitedefaultmidpunct}
{\mcitedefaultendpunct}{\mcitedefaultseppunct}\relax
\EndOfBibitem
\bibitem[Schmidt and Lemeshko(2015)Schmidt, and Lemeshko]{lemesheko:2015:prl}
Schmidt,~R.; Lemeshko,~M. Rotation of Quantum Impurities in the Presence of a Many-Body Environment. \emph{Physical Review Letters} \textbf{2015}, \emph{114}, 203001\relax
\mciteBstWouldAddEndPuncttrue
\mciteSetBstMidEndSepPunct{\mcitedefaultmidpunct}
{\mcitedefaultendpunct}{\mcitedefaultseppunct}\relax
\EndOfBibitem
\end{mcitethebibliography}
\end{document}